\documentclass[aps,pra,a4paper,reprint,superscriptaddress,amsmath,amssymb]{revtex4-1}
\usepackage{bbm}
\usepackage[utf8]{inputenc}

\newcommand{\mr}{\mathrm}               
\newcommand{\mf}{\mathbf}               
\newcommand{\ms}{\mathsf}               

\newcommand{\<}{\langle}                
\renewcommand{\>}{\rangle}              
\newcommand{\pd}{\partial}              
\renewcommand{\d}{\mathrm{d}}           
\renewcommand{\dag}{^{\dagger}}         

\newcommand{\cA}{\mathcal{A}}           
\newcommand{\cM}{\mathcal{M}}           
\newcommand{\UI}{\hat{U}_I(t,t_0)}      

\renewcommand{\H}{\hat{H}}              
\newcommand{\HH}{\hat{\mathcal{H}}}     
\renewcommand{\aa}{\hat{a}}             
\newcommand{\cc}{\hat{c}}               
\newcommand{\uu}{\hat{u}}               
\renewcommand{\AA}{\hat{\mathbf{A}}}    
\newcommand{\JJ}{\hat{\mathbf{J}}}      
\newcommand{\rr}{\hat{\mathbf{r}}}      
\newcommand{\pp}{\hat{\mathbf{p}}}      
\newcommand{\RR}{\hat{\mathbf{R}}}      
\newcommand{\PP}{\hat{\mathbf{P}}}      
\newcommand{\rrho}{\hat{\rho}}          
\newcommand{\Rset}{\{\mathbf{R}_I\}}    
\newcommand{\RRset}{\{\hat{\mathbf{R}}_I\}} 
\newcommand{\Roset}{\{\mathbf{R}^{(0)}_I\}} 
\newcommand{\ppsi}{\hat{\psi}}          

\renewcommand{\k}{\mathbf{k}}           
\newcommand{\kin}{\mathbf{k}_{\mathrm{in}}} 
\newcommand{\kout}{\mathbf{k}_{\mathrm{out}}} 
\renewcommand{\r}{\mathbf{r}}           
\newcommand{\R}{\mathbf{R}}             
\newcommand{\q}{\mathbf{q}}             

\renewcommand{\a}{\alpha}               
\renewcommand{\b}{\beta}                
\newcommand{\g}{\gamma}                 
\newcommand{\G}{\Gamma}                 
\newcommand{\e}{\varepsilon}            
\renewcommand{\l}{\lambda}              
\newcommand{\m}{\mu}                    
\newcommand{\n}{\nu}                    
\renewcommand{\o}{\omega}               
\newcommand{\oin}{\omega_{\mathrm{in}}} 
\newcommand{\oout}{\omega_{\mathrm{out}}} 
\newcommand{\oL}{\omega_{\mathrm{L}}}   
\newcommand{\oD}{\omega_{\mathrm{D}}}   

\newcommand{\ad}{(\mathrm{ad.})}        
\newcommand{\he}{\hat{h}_{\mathrm{e}}}  
\newcommand{\Eeo}{E^{(\mathrm{ad.})}_{\mathrm{e},0}} 
\newcommand{\0}{\emptyset}              

\begin{document}


\title{ Theory of resonant Raman scattering: Toward a comprehensive \textit{ab initio} description }

\author{Sven Reichardt}
\affiliation{Department of Materials, University of Oxford, Parks Road, Oxford OX1 3PH, United Kingdom}
\affiliation{Physics and Materials Science Research Unit, University of Luxembourg, 1511 Luxembourg, Luxembourg}

\author{Ludger Wirtz}
\affiliation{Physics and Materials Science Research Unit, University of Luxembourg, 1511 Luxembourg, Luxembourg}


\begin{abstract}

We develop a general, fully quantum mechanical theory of Raman scattering from first principles
in terms of many-body correlation functions.
In order to arrive at expressions that are practically useful in the context of condensed matter physics,
we adopt the Lehmann-Symanzik-Zimmermann reduction formula from high-energy physics
and formulate it in the language of many-body perturbation theory.
This enables us to derive a general and practically useful expression for the Raman
scattering rate in terms of quantities that can be computed \textit{ab initio}.
Our work paves the way toward a comprehensive computational approach to the calculation of Raman spectra
that goes beyond the current state of the art by capturing both excitonic and non-adiabatic effects.

\end{abstract}

\maketitle


\section{Introduction}
\label{sec:intro}

Raman spectroscopy has developed into one of the most reliable tools for the characterization of materials.
Its ability to probe both electronic and vibrational properties at once has made it a popular tool for the investigation
of materials, especially low-dimensional ones, such as graphene or transition metal dicalchogenides.
Amongst others, it allows the probing of the number of layers~\cite{ferrari2006,graf2007},
strain fluctuations on different length scales down to the nm-scale
~\cite{mohiuddin2009,huang2010,mohr2010,yoon2011,lee2012,zabel2012,chacon2013,engels2014b,neumann2015b},
the amount, nature, and distribution of doping and doping domains
~\cite{lee2012,ferrari2007,yan2007,pisana2007,stampfer2007},
the lattice temperature~\cite{calizo2007,balandin2008},
many-body effects such as electron and phonon lifetimes
~\cite{faugeras2009,yan2010,faugeras2011,faugeras2012,kossacki2012,goler2012,kim2013,qiu2013,
leszczynski2014,berciaud2014,neumann2015,neumann2015c,faugeras2015,sonntag2018},
and excitonic effects~\cite{miranda2017}.
This versatility, however, makes a theoretical description very challenging and so far,
no complete and fully general theory from first principles exists.
In particular, there does not yet exist a coherent, fully quantum-mechanical description that
takes into account both the effects of electron-electron interaction, such as exciton formation,
and electron-phonon interaction beyond the static or adiabatic limit.
Both of the latter affect both the height, width, and shape of the peaks present in the Raman spectrum.

The modern, \textit{ab initio} approaches to the calculation of Raman spectra focus almost exclusively
on processes with either one or two phonons in the final state and the calculation of the corresponding scattering rates,
i.e., the intensity of the corresponding peaks in the spectrum of the scattered light.
Among these approaches are, in increasing order of complexity and accuracy:
a semi-empirical bond polarizability model, in which the polarizability of the inter-atomic chemical bonds
is parametrized and fitted to experiment~\cite{saito1998b,umari2001,wirtz2005,luo2013,luo2013b,froehlicher2015};
an approach based on density functional perturbation theory for the calculation of the mixed third derivative
of the total ground state energy with respect to two external electric fields
and an atomic displacement or lattice distortion~\cite{lazzeri2003,veithen2005};
the computation of the quantum mechanical scattering matrix element
in third- or fourth-order perturbation theory~\cite{basko2008,basko2009,venezuela2011,herziger2014,hasdeo2016,reichardt2017b};
and the computation of the first derivative of the dielectric susceptibility with respect to static atomic displacements
via the method of finite differences~\cite{knoll1995,ambrosch2002,gillet2013,miranda2017,gillet2017b}.
However, each of these methods captures only certain aspects of the Raman process while being insensitive to others.
Finite-difference methods~\cite{knoll1995,ambrosch2002,gillet2013,miranda2017,gillet2017b}, for example,
allow the inclusion of electron-electron/-hole interactions, i.e., excitonic effects,
but treat the nuclear and vibrational degrees of freedom entirely statically.
The existing perturbative methods~\cite{basko2008,basko2009,venezuela2011,herziger2014,hasdeo2016,reichardt2017b},
on the other hand, can capture the dynamical effects of nuclear vibrations, but cannot capture excitonic effects.
Up to now there is no comprehensive method available that both incorporates the effects of electron-electron/-hole interaction
and allows the inclusion of dynamical and non-adiabatic effects of nuclei vibrations.
The combination of both can be important both in low-dimensional systems with small electronic band gaps,
such as carbon nanotubes, or for materials whose properties are \emph{a priori} unknown, as encountered in high-throughput materials screening.
As such, a comprehensive method for the calculation of Raman spectra from first principles
that does not focus on describing only a certain aspect of the Raman scattering process is highly desirable.

Here we aim to fill this gap by developing a general formulation of inelastic light scattering
starting from the fundamental quantum mechanical many-body Hamiltonian.
The main goal of this work is to lay the theoretical foundations
for a practical \textit{ab initio} approach to the calculation of Raman spectra.
To this end, we first demonstrate how the probability for inelastic light scattering
within a finite time interval and at finite temperature can be expressed entirely in terms of
a correlation of quantities intrinsic to an interacting system of electrons and nuclei.
To arrive at a general and practically useful expression, we then specialize
to the case of low temperatures and long time intervals and derive a generalized version of Fermi's golden rule.
Finally, we show how applying a well-known concept from high-energy physics,
the Lehmann-Symanzik-Zimmermann (LSZ) reduction formula~\cite{lehmann1955},
in a condensed matter physics context allows the formulation of an expression for the Raman scattering rate
entirely in terms of time-ordered correlation and spectral functions.
The latter formulation in the modern language of many-body perturbation theory
opens the door to a timely computational realization of our suggested approach~\cite{reichardt2018,reichardt2018c}.


\section{Light-matter interaction}
\label{sec:matlight}

The starting point of this work is the total Hamiltonian of an interacting system of light and matter:
\begin{equation}
  \HH = \HH_{\mr{Light}} + \HH_{\mr{Light-Matter}} + \HH_{\mr{Matter}}.
  \label{eq:htot}
\end{equation}
We will treat both the light and the matter system in a fully quantum mechanical way,
with operators in calligraphic font acting in the Hilbert space of both light and matter.
The three pieces of the total Hamiltonian in Eq.~(\ref{eq:htot}) describe, in order, a system of non-interacting photons,
the interaction of photons with matter, and the isolated matter system.

The free light Hamiltonian, effectively acting only in the Hilbert space of the photons,
has the form $\HH_{\mr{Light}} = \H_{\mr{L}}\otimes\hat{\mathbbm{1}}_{\mr{M}}$.
The Schrödinger-picture Hamiltonian $\H_{\mr{L}}$ is the normal-ordered operator corresponding
to the classical Hamilton function for electromagnetic fields, which, in Gaussian units, is given by~\cite{jackson2007}
\begin{equation}
  H_{\mr{L}} \equiv \int \d^3 r \, \frac{1}{8\pi} \bigg[ \Big(\mf{E}(\r,t)\Big)^2 + \Big(\mf{B}(\r,t)\Big)^2 \bigg].
\end{equation}
The electric and magnetic fields can be written in terms of the scalar and vector potentials as
\begin{equation}
  \mf{E}(\r,t) = -\frac{1}{c}\frac{\pd}{\pd t}\mf{A}(\r,t) -\nabla \phi(\r,t),
  \quad \mf{B}(\r,t) = \nabla \times \mf{A}(\r,t).
\end{equation}
We will work in Coulomb or \emph{radiation} gauge, i.e., ${\nabla \cdot \mf{A}(\r,t) = 0}$, in which Maxwells's equations are equivalant to
\begin{equation}
  \nabla^2 \phi(\r,t) = 0, \ \left(\nabla^2 - \frac{1}{c^2}\frac{\pd^2}{\pd t^2}\right) \mf{A}(\r,t) = \frac{1}{c}\frac{\pd}{\pd t} \nabla \phi(\r,t).
\end{equation}
Demanding that the scalar potential vanish for $\r \to \infty$ leads to the unique solution $\phi(\r,t)\equiv0$ for Laplace's equation.
The remaining wave equation for the vector potential is solved by expanding $\mf{A}$ in terms of plane waves.
For convenience, we treat the system as being embedded in a finite, but large volume $V$ with the vector potential vanishing on the boundary.
Upon quantization, the Schrödinger-picture operator representing the vector potential (in units with $\hbar\equiv1$) reads
\begin{equation}
 \AA(\r) = \sum_{\k,\m} \sqrt{\frac{4 \pi c^2}{2 \o_{\k} V}} \left(  \aa_{\k,\m} \boldsymbol{\epsilon}_{\k,\m} \mr{e}^{i\k\cdot\r}
                                                                   + \aa\dag_{\k,\m} \boldsymbol{\epsilon}^*_{\k,\m} \mr{e}^{-i\k\cdot\r} \right).
\end{equation}
The sum runs over all wave vectors $\k$ that are compatible with the boundary condition $\mf{A}(\r)|_{\pd V} = \mf{0}$
and over the two possible polarizations labeled by $\m$ and described by the two mutually orthogonal polarization vectors
$\boldsymbol{\epsilon}_{\k,\m=1,2}$ that also satisfy the Coulomb gauge condition $\k \cdot \boldsymbol{\epsilon}_{\k,\m=1,2}=0$.
The annihilation and creation operators $\aa_{\k,\m}$ and $\aa\dag_{\k,\m}$ obey the usual bosonic commutation relations~\cite{schwabl2005}.
In terms of them and the light frequency $\o_{\k} \equiv c|\k|$, the Hamiltonian for a non-interacting system of photons
\footnote{Note that within the formalism of this work, any modification of the light dispersion by the matter system,
e.g., a non-unit refractive index in solids, will appear as a consequence of the light-matter interaction
and can be incorporated within the framework of perturbation theory by computing the photon self-energy.
However, since we will not make explicit use of the precise form of the light dispersion, this point will not be discussed any further.} reads
\begin{equation}
  \H_{\mr{L}} = \sum_{\k,\m} \o_{\k} \aa\dag_{\k,\m} \aa_{\k,\m}.
\end{equation}

Next, we turn to the description of the non-interacting matter system.
The corresponding Hamiltonian reads $\HH_{\mr{Matter}} = \hat{\mathbbm{1}}_{\mr{L}}\otimes\H_{\mr{M}}$,
where the operator $\H_{\mr{M}}$ acts only in the matter Hilbert space.
The most general form describing a system of non-relativistic electrons and nuclei mutually interacting with each other via Coulomb interaction is
\begin{widetext}
  \begin{equation}
    \H_{\mr{M}} =   \sum_i \frac{\pp^2_i}{2 m} + \frac{1}{2}\sum_{\substack{i,j \\i \neq j}} \frac{e^2}{|\rr_i-\rr_j|}
                  + \sum_I \frac{\PP^2_I}{2 M_I} + \frac{1}{2}\sum_{\substack{I,J \\I \neq J}} \frac{Z_I Z_J e^2}{|\RR_I-\RR_J|}
                  + \sum_{i,I} \frac{-Z_Ie^2}{|\rr_i-\RR_I|}.
    \label{eq:hmatter}
  \end{equation}
\end{widetext}
Here, lower case (upper case) operators and indices correspond to electrons (nuclei).
The mass of the electrons and the masses of the nuclei are denoted by $m$ and $M_I$, respectively,
while the electron and nuclei charges are given by $-e<0$ and $Z_I e>0$, respectively.
We want to point out that this Hamiltonian is completely general and treats the electrons and nuclei on the same quantum mechanical footing.

Finally, we introduce a gauge-invariant coupling between the matter and light-systems via the minimal coupling prescription~\cite{cohen1977}
\begin{equation}
  \pp_i \to \pp_i + \frac{e}{c} \AA(\rr_i), \qquad \PP_I \to \PP_I - \frac{Z_I e}{c} \AA(\RR_I).
\end{equation}
Note that the vector potential operator $\AA(\r)$ acts in the Hilbert space of the photons only,
while the position and momentum operators act in the Hilbert space of the matter system.
The operator $\AA(\rr)$, on the other hand, acts in the full Hilbert space as it
contains a product of both photon and matter operators in the form of $\aa_{\k,\m} \exp(i\k\cdot\r)$ and its hermitian conjugate
\footnote{The notation $\aa_{\k,\m} \exp(i\k\cdot\rr)$ is understood to be a short-hand notation for $\aa_{\k,\m} \otimes \exp(i\k\cdot\rr)$
\label{fnote:tensorprod}}.
Expanding the squares of the kinetic momentum operators in the matter Hamiltonian yields
\begin{equation}
  \begin{split}
    \H_{\mr{M}} \to \H_{\mr{M}} &+ \sum_i \frac{e}{m c} \AA(\rr_i) \cdot \pp_i - \sum_I \frac{Z_I e}{M_I c} \AA(\RR_I) \cdot \PP_I \\
                                &+ \sum_i \frac{e^2}{2 m c^2} \AA^2(\rr_i) + \sum_I \frac{Z_I^2 e^2}{2 M_I c^2} \AA^2(\RR_I).
  \end{split}
\end{equation}
where we made use of the Coulomb gauge condition $\nabla \cdot \AA(\r)=0$ to combine the terms of the form $\AA(\rr)\cdot\pp + \pp \cdot \AA(\rr)$,
which involve both $\AA(\rr)$ and $\pp$, operators which would normally not commute.

The light-matter interaction Hamiltonian can be written in a more familiar form
by defining the total matter current density operator as
\begin{equation}
  \JJ(\r) \equiv \sum_i (-e)\delta^{(3)}(\r-\rr_i)\frac{\pp_i}{m} + \sum_I (Z_I e)\delta^{(3)}(\r-\RR_I)\frac{\PP_I}{M_I},
  \label{eq:current}
\end{equation}
where $\delta^{(3)}(\r)$ denotes the three-dimensional Dirac $\delta$-distribution
\footnote{
Note that, in general, the operators $\delta^{(3)}(\r-\rr_i)$ and $\pp_i$ do not commute and that, in principle, a correct passing
from the classical expression to the quantum mechanical one would require the invoking of Weyl's symmetrization postulate
$f(\r)g(\mf{p})\to[f(\rr)g(\pp)+g(\pp)f(\rr)]/2$.
The application of this postulate to the current density would lead to the familiar form of the probability current times the electric charge,
when taking the expectation value of $\JJ(\r)$ in a state {$|\psi\>$}: $\hbar/(2mi)[\psi^*(\r)\nabla\psi(\r) - \psi(\r)\nabla\psi^*(\r)]$.
In Coulomb gauge, however, an integration by parts of the second term reduces this expression to the non-symmetrized one of Eq.~(\ref{eq:current}).
We have already made use of this argument in the derivation of the light-matter Hamiltonian,
by using the fact that $\AA(\rr) \cdot \pp = \pp \cdot \AA(\rr)$ in Coulomb gauge
\label{fnote:rpordering}}.
Note that each term in it has the schematic form $\mf{J}(\r)\sim\varrho(\r-\r_0)\mf{v}$, where $\varrho(\r-\r_0)$
represents the charge density of a point particle at position $\r_0$ and $\mf{v}$ its velocity,
as familiar from the classical electrodynamics of point particles~\cite{jackson2007}.
In terms of the total matter current density operator and the vector potential,
the Hamiltonian for the interaction between light and matter
in the approximation of neglecting the $\AA^2$-terms reads:
\begin{equation}
  \HH_{\mr{Light-Matter}} = -\frac{1}{c}\int \d^3 r \, \mf{\AA}(\r) \cdot \JJ(\r).
  \label{eq:hlightmatter}
\end{equation}
We want to stress that, in the equation above, $\AA(\r)$ now acts only on the Hilbert space of photons,
while $\JJ(\r)$ is entirely restricted to the matter Hilbert space~
\footnote{We use the short-hand notation ${\mf{\AA}(\r) \cdot \JJ(\r) \equiv \sum_i \hat{A}_i(\r) \otimes \hat{J}_i(\r)}$
(compare remark in~\cite{Note1})}.
Having specified the different terms in the total Hamiltonian $\HH$ as given in Eq.~(\ref{eq:htot}),
we will now discuss the treatment of inelastic light scattering by the matter system within the framework of perturbation theory.


\section{Correlation-function formulation}
\label{sec:corfun}

Our approach to the description of inelastic light scattering follows the most commonly encountered experimental setting nowadays.
We start at time $t_0$ with a matter sample that can potentially be in contact with a heat bath at a finite temperature
and an incoming photon with wave vector $\kin$ and polarization $\m$.
We are interested in the probability that the light-matter system at a later time $t$ is in a state with
one outgoing photon of different wave vector $\kout$ and polarization $\n$ with the matter system being in an arbitrary state $|\a\>$.

Mathematically, this process is most conveniently described in the language of the density matrix $\rrho(t)$,
whose time-evolution is governed by the von Neumann equation~\cite{cohen1977}
\begin{equation}
  i\frac{\pd}{\pd t} \rrho(t) = [\HH,\rrho(t)].
\end{equation}
As the total Hamiltonian $\HH$ is time-independent, its solution is given by
\begin{equation}
  \rrho(t) = \mr{e}^{-i\HH(t-t_0)} \rrho(t_0) \mr{e}^{+i\HH(t-t_0)},
\end{equation}
where $\rho(t_0)$ is the density matrix at an initial time $t_0$,
\begin{equation}
  \rrho(t_0) = |\kin,\m\>\<\kin,\m| \otimes \frac{1}{Z_{\mr{M}}} \mr{e}^{-\b \H_{\mr{M}}},
\end{equation}
which describes the photon system being in a one-photon state $|\kin,\m\>$
and the matter system being in thermal equilibrium with a heat bath at temperature $T=(k_B \b)^{-1}$.
The normalization of $\rrho(t_0)$ is ensured by letting $Z_{\mr{M}} = \mr{tr} \exp(-\b \H_{\mr{M}})$,
where $\H_{\mr{M}}$ is the Hamiltonian of the matter system, given in Eq.~(\ref{eq:hmatter}).

The probability for the light-matter system to be in a different one-photon state $|\kout,\nu\>$ and an arbitrary matter state
$|\a\>$ at time $t$ is given by a partial sum of diagonal elements of the density matrix at time $t$:
\begin{equation}
  P_{\mr{scatter}} = \sum_{\a} \Big( \<\kout,\n| \otimes \<\a| \Big) \rrho(t) \Big( |\kout,\n\> \otimes |\a\> \Big),
\end{equation}
where the sum over $\a$ runs over a complete set of matter states.
For later convenience, we choose this set as coinciding with the complete set of eigenstates
of the matter Hamiltonian $\H_{\mr{M}}$, i.e., $\H_{\mr{M}} |\a\> = E_{\a} |\a\>$.
By expressing the matter part of the initial density matrix in this basis,
\begin{equation}
  \frac{1}{Z_{\mr{M}}} \mr{e}^{-\b \H_{\mr{M}}} = \frac{1}{Z_{\mr{M}}} \sum_{\g} \mr{e}^{-\b E_{\g}} |\g\>\<\g|,
\end{equation}
the scattering probability reads
\begin{widetext}
  \begin{equation}
    P_{\mr{scatter}} = \frac{1}{Z_{\mr{M}}}\sum_{\a,\g} \mr{e}^{-\b E_{\g}} \bigg| \Big( \<\kout,\n| \otimes \<\a| \Big)
                       \mr{e}^{-i \HH (t-t_0)} \Big( |\kin,\m\> \otimes |\g\> \Big) \bigg|^2.
    \label{eq:prob-fermi}
  \end{equation}
\end{widetext}

In order to evaluate the needed matrix elements of the time-evolution operator $\exp(-i\HH(t-t_0))$,
we employ the formalism of time-dependent perturbation theory in the interaction picture~\cite{schwabl2005}.
To start with, we define the time-evolution operator in the interaction picture:
\begin{equation}
  \UI \equiv \mr{e}^{+i \HH_0 t} \mr{e}^{-i \HH (t-t_0)} \mr{e}^{-i \HH_0 t_0}.
\end{equation}
In terms of $\UI$, the needed matrix elements read:
\begin{equation}
  \begin{split}
    &  \Big( \<\kout,\n| \otimes \<\a| \Big) \mr{e}^{-i \HH (t-t_0)} \Big( |\kin,\m\> \otimes |\g\> \Big) \\
    & \qquad =  \mr{e}^{-i(E_{\a}+\oout)t} \mr{e}^{+i(E_{\g}+\oin)t_0} \\
    & \qquad \quad \times \Big( \<\kout,\n| \otimes \<\a| \Big) \UI \Big( |\kin,\m\> \otimes |\g\> \Big),
  \end{split}
  \label{eq:me-inter}
\end{equation}
where $\oin \equiv \o_{\kin}, \oout \equiv \o_{\kout}$ and the oscillating exponential prefactors are inconsequential
as they drop out after taking the modulus squared of the matrix elements.
To evaluate the matrix elements of $\UI$, we use its exponential representation for $t \geq t_0$
~\cite{abrikosov1965,fetter1971,peskin1995,mahan2000,schwabl2005,weinberg2005},
\begin{equation}
  \UI = \mathcal{T} \exp\left\{ -i \int_{t_0}^t \d t' \, \HH_{1,I}(t') \right\},
  \label{eq:ui}
\end{equation}
which obeys the equation of motion
\begin{equation}
  i \frac{\pd}{\pd t} \UI = \HH_{1,I}(t) \UI.
\end{equation}
Here, $\mathcal{T}$ is the time-ordering symbol and we defined the interaction Hamiltonian in the interaction picture as
\begin{equation}
  \HH_{1,I}(t) \equiv \mr{e}^{i \HH_0 t} \HH_{\mr{Light-Matter}} \mr{e}^{-i \HH_0 t},
\end{equation}
with $\HH_0 \equiv \HH_{\mr{Light}} + \HH_{\mr{Matter}}$.
The exponential representation of Eq.~(\ref{eq:ui}) permits a convenient perturbative treatment of the light-matter interaction
via a Taylor expansion of the exponential.

The description of inelastic light scattering in the formalism presented here requires the knowledge of
the matrix elements of $\UI$ that involve one-photon states with different wave vectors.
Since $\HH_{1,I}(t)$ is linear in the vector potential and thus creates or destroys one photon,
the lowest-order, non-vanishing contribution to the relevant matrix elements is given by the second-order term
of the Taylor series of $\UI$:
\begin{equation}
  \begin{split}
          & \Big( \<\kout,\n| \otimes \<\a| \Big) \UI \Big( |\kin,\m\> \otimes |\g\> \Big) \\
    \simeq& \sum_{i,j} \int_{t_0}^t \d t_1 \int_{t_0}^t \d t_2 \int \d^3 r_1 \int \d^3 r_2 \, \frac{(-i)^2}{2! c^2} \\
          & \times \<\a| \mathcal{T}\left[ \hat{J}_{i,I}(\r_1,t_1) \hat{J}_{j,I}(\r_2,t_2) \right] |\g\> \\
          & \times \<\kout,\n| \mathcal{T}\left[ \hat{A}_{i,I}(\r_1,t_1) \hat{A}_{j,I}(\r_2,t_2) \right] |\kin,\m\>,
  \end{split}
  \label{eq:ui-me}
\end{equation}
where the sums over $i,j$ run over the three Cartesian components of the vector operators.
The main benefit of the perturbative treatment is the factorization of the needed matrix elements
into a product of separate matrix elements in the matter and light Hilbert spaces,
the latter of which is straightforward to evaluate by using Wick's theorem
~\cite{abrikosov1965,fetter1971,peskin1995,mahan2000,schwabl2005,weinberg2005}:
\begin{widetext}
  \begin{equation}
    \begin{split}
         \<\kout,\n| \mathcal{T}\left[ \hat{A}_{i,I}(\r_1,t_1) \hat{A}_{j,I}(\r_2,t_2) \right] |\kin,\m\>
      =& \ \delta_{\kin,\kout} \delta_{\m,\n} \<0_{\mr{L}}| \mathcal{T}\left[ \hat{A}_{i,I}(\r_1,t_1) \hat{A}_{j,I}(\r_2,t_2) \right] |0_{\mr{L}}\> \\
       & + \frac{2 \pi c^2}{\sqrt{\oin \oout}V} \mr{e}^{-i(\kout\cdot\r_1 - \oout t_1)}
           \mr{e}^{+i(\kin\cdot\r_2 - \oin t_2)} \Big(\epsilon^i_{\kout,\n}\Big)^* \epsilon^j_{\kin,\m} \\
       & + \frac{2 \pi c^2}{\sqrt{\oin \oout}V} \mr{e}^{-i(\kout\cdot\r_2 - \oout t_2)}
           \mr{e}^{+i(\kin\cdot\r_1 - \oin t_1)} \Big(\epsilon^j_{\kout,\n}\Big)^* \epsilon^i_{\kin,\m}.
    \end{split}
  \end{equation}
\end{widetext}
While the first term does not contribute to inelastic light scattering and hence will be dropped in the following,
the second and third terms differ only by the exchange $(\r_1,t_1,i) \leftrightarrow (\r_2,t_2,j)$
and hence give the same contribution to the matrix elements of $\UI$ as the integrations
and summations in Eq.~(\ref{eq:ui-me}) are symmetric under the exchange of integration variables and summation indices.
We can simplify the notation by introducing the spatially Fourier-transformed and projected current operators via
\begin{equation}
  \hat{J}_{\k,\m,I}(t) \equiv \boldsymbol{\epsilon}^*_{\k,\m} \cdot \int \d^3 r \, \mr{e}^{-i\k\cdot\r} \JJ_I(\r,t).
  \label{eq:currentfour}
\end{equation}
In terms of these operators, the matrix elements of $\UI$ take on the form
\begin{widetext}
  \begin{equation}
           \Big( \<\kout,\n| \otimes \<\a| \Big) \UI \Big( |\kin,\m\> \otimes |\g\> \Big)
    \simeq \frac{-2\pi}{V \sqrt{\oin \oout}} \int_{t_0}^t \d t_1 \int_{t_0}^t \d t_2 \,
           \mr{e}^{i(\oout t_1 - \oin t_2)} \<\a| \mathcal{T}\left[ \hat{J}_{\kout,\n,I}(t_1) \hat{J}\dag_{\kin,\m,I}(t_2) \right] |\g\>.
    \label{eq:ui-me-cur}
  \end{equation}
\end{widetext}
Combining Eqs.~(\ref{eq:prob-fermi}), (\ref{eq:me-inter}), and (\ref{eq:ui-me-cur}) and using the completeness relation for the matter states,
$\sum_{\a} |\a\>\<\a| = \mathbbm{1}_{\mr{M}}$, we arrive at a compact expression for the probability for inelastic light scattering:
\begin{widetext}
  \begin{equation}
    \begin{split}
      P_{\mr{inel.}} \simeq& \frac{(2\pi)^2}{V^2 \oin \oout}
                             \int_{t_0}^t \d t_1 \int_{t_0}^t \d t_2 \int_{t_0}^t \d t'_1 \int_{t_0}^t \d t'_2 \,
                             \mr{e}^{i\oout(t_1-t'_1)} \mr{e}^{-i\oin(t_2-t'_2)} \\
                           & \times \left\< \overline{\mathcal{T}} \left[ \hat{J}\dag_{\kout,\n,I}(t'_1) \hat{J}_{\kin,\m,I}(t'_2) \right]
                             \mathcal{T} \left[ \hat{J}_{\kout,\n,I}(t_1) \hat{J}\dag_{\kin,\m,I}(t_2) \right] \right\>_{\mr{M}},
    \end{split}
  \end{equation}
\end{widetext}
where $\overline{\mathcal{T}}$ denotes the anti-time-ordering symbol,
which arises due to the reversal of the order of the operators under complex conjugation.
We also identified the thermal and quantum mechanical expectation value of an operator
acting in the matter Hilbert space: $\<\hat{O}\>_{\mr{M}} \equiv Z^{-1}_{\mr{M}} \mr{tr}\,[\exp(-\b\H_{\mr{M}})\hat{O}]$.

This expression gives the probability for one photon with momentum $\kin$ and polarization $\m$ to scatter
inelastically to a state $|\kout,\n\> \neq |\kin,\m\>$ when it interacts with a matter system over a time period $t-t_0$.
In an experimental setting, however, one cannot detect a photon with a precise momentum.
Instead, a detector always detects all scattered photons within a certain direction in a small solid angle
$\Delta\Omega_{\mr{D}}$ and within a small, but finite frequency interval $[\oD,\oD + \Delta\oD]$.
It is thus more sensible to consider the total probability for all such processes.
Similarly, the source of the incoming photons typically emits photons into a very small, but finite
solid angle $\Delta\Omega_{\mr{L}}$ over a finite frequency interval $[\oL,\oL + \Delta\oL]$.
If the $\Delta\Omega_{\mr{D,L}}$ and $\Delta\o_{\mr{D,L}}$ are small enough,
the total scattering probability can be approximated by its value evaluated at $\oin \equiv c|\kin| = \oL$ and $\oout \equiv c|\kout|= \oD$
and in the direction of $\kin$ and $\kout$, specified by the axis of the incoming light and the position of the detector, respectively
and multiplied by the number of photon states in this frequency interval and solid angle range:
\begin{equation}
  \begin{split}
       N_{\text{photon states}}(c|\k|\in[\o,\o+\Delta\o], &\k/|\k| \in \Delta\Omega) \\
    =& \frac{V \o^2}{(2\pi)^3 c^3} \, \Delta\Omega \, \Delta\o.
  \end{split}
\end{equation}
Including this kinematic factor for both the incoming and outgoing light then yields the total scattering probability:
\begin{widetext}
  \begin{equation}
    \begin{split}
      P_{\mr{inel.}} \simeq& \frac{\oL\oD \Delta\Omega_{\mr{L}} \Delta\oL \Delta\Omega_{\mr{D}} \Delta\oD}{(2\pi)^4 c^6}
                             \int_{t_0}^t \d t_1 \int_{t_0}^t \d t_2 \int_{t_0}^t \d t'_1 \int_{t_0}^t \d t'_2 \,
                             \mr{e}^{i\oD(t_1-t'_1)} \mr{e}^{-i\oL(t_2-t'_2)} \\
                           & \times \left\< \overline{\mathcal{T}} \left[ \hat{J}\dag_{\kout,\n,I}(t'_1) \hat{J}_{\kin,\m,I}(t'_2) \right]
                             \mathcal{T} \left[ \hat{J}_{\kout,\n,I}(t_1) \hat{J}\dag_{\kin,\m,I}(t_2) \right] \right\>_{\mr{M}}.
      \label{eq:corfun}
    \end{split}
  \end{equation}
\end{widetext}

Eq.~(\ref{eq:corfun}) is one of the central results of this work.
It expresses the probability for inelastic light scattering in a \emph{finite} time interval in terms of a thermal correlation function
of operators acting in the matter part of the Hilbert space only.
As such, it provides a way to calculate the intensity of inelastically scattered light on \emph{arbitrarily short time scales},
which is a first step towards a theoretical description of ultra-fast Raman spectroscopy.
It also does not make any use of specific intermediate or final states of the matter system
and hence it provides the \emph{complete} Raman scattering probability, including \emph{all possible matter excitations},
whereas the case commonly discussed in the literature focuses on one specific excitation only, for instance, vibrations of the nuclei.

While a more detailed discussion of this correlation could potentially be done within the Keldysh contour formalism, this is beyond the scope of work.
Instead, we focus on recasting the elegant, yet abstract correlation-function formulation into a more practically useful form
by deriving a generalized Fermi's golden rule that allows the consideration of individual contributions to the scattering process
in perturbation theory.
This formulation is especially suitable for a computational implementation.


\section{Generalized Fermi's golden rule}
\label{sec:fermi}

To derive a generalized Fermi's golden rule for inelastic light scattering, we start from Eq.~(\ref{eq:prob-fermi})
and approximate the matrix element of the time-evolution operator in the interaction picture as in Eq.~(\ref{eq:ui-me-cur}).
This leads to
\begin{widetext}
  \begin{equation}
    P_{\mr{inel.}} \simeq \G_{\mr{kin.}}(\oL,\oD) \frac{1}{Z_{\mr{M}}} \sum_{\a,\g} \mr{e}^{-\b E_{\g}}
                          \left| \int_{t_0}^t \d t_1 \int_{t_0}^t \d t_2\, \mr{e}^{i(\oD t_1 - \oL t_2)}
                          \<\a| \mathcal{T}\left[ \hat{J}_{\kout,\n,I}(t_1) \hat{J}\dag_{\kin,\m,I}(t_2) \right] |\g\> \right|^2,
    \label{eq:p-inel}
  \end{equation}
\end{widetext}
where
\begin{equation}
  \G_{\mr{kin.}}(\oL,\oD) \equiv \frac{\oL\oD \Delta\Omega_{\mr{L}} \Delta\oL \Delta\Omega_{\mr{D}} \Delta\oD}{(2\pi)^4 c^6}
  \label{eq:gam-kin}
\end{equation}
is the kinematic (phase space) factor introduced before.
Next, we note that the matrix element of the time-ordered product of two current operators has the form
\begin{widetext}
  \begin{equation}
    \begin{split}
        \<\a| \mathcal{T}\left[ \hat{J}_{\kout,\n,I}(t_1) \hat{J}\dag_{\kin,\m,I}(t_2) \right] |\g\>
      = \mr{e}^{i(E_{\a}-E_{\g})t_2} \Big[ &  \mr{e}^{i E_{\a} (t_1-t_2)} \theta(t_1-t_2)
                                              \<\a|\hat{J}_{\kout,\n}\mr{e}^{-i\H_{\mr{M}}(t_1-t_2)}\hat{J}\dag_{\kin,\m}|\g\> \\
                                           &+ \mr{e}^{-i E_{\g} (t_1-t_2)} \theta(-(t_1-t_2))
                                              \<\a|\hat{J}_{\kin,\m}\mr{e}^{i\H_{\mr{M}}(t_1-t_2)}\hat{J}\dag_{\kout,\n}|\g\> \Big].
    \end{split}
  \end{equation}
\end{widetext}
Barring the first exponential factor, it is a function of the time difference $t_1-t_2$ only.
We can then introduce its Fourier decomposition in the form
\begin{equation}
  \begin{split}
     & \ \<\a| \mathcal{T}\left[ \hat{J}_{\kout,\n,I}(t_1) \hat{J}\dag_{\kin,\m,I}(t_2) \right] |\g\> \\
    =& \ \mr{e}^{i(E_{\a}-E_{\g})t_2} \int \frac{\d \o}{2 \pi} \, \mr{e}^{-i \o (t_1-t_2)}\tilde{J}_{\a\g}(\o;\Lambda).
  \end{split}
\end{equation}
Here, the Fourier components are defined as
\begin{equation}
  \tilde{J}_{\a\g}(\o;\Lambda)
  \equiv \int_{-\infty}^{+\infty} \d t \, \mr{e}^{i \o t} \<\a| \mathcal{T}\left[ \hat{J}_{\kout,\n,I}(t) \hat{J}\dag_{\kin,\m,I}(0) \right] |\g\>,
  \label{eq:me-freq}
\end{equation}
where $\Lambda\equiv(\kout,\nu;\kin,\m)$ is a short-hand notation introduced to avoid a cluttering of the expression with too many indices.
Using the Fourier decomposition of the matrix elements, the integrations in the time domain in Eq.~(\ref{eq:p-inel}) are straightforward and yield
\begin{widetext}
  \begin{equation}
    \begin{split}
      P_{\mr{inel.}} \simeq& \ \G_{\mr{kin.}}(\oL,\oD) (t-t_0)^4 \frac{1}{Z_{\mr{M}}} \sum_{\a,\g} \mr{e}^{-\b E_{\g}}
                             \int \frac{\d \o}{2 \pi} \int \frac{\d \o'}{2 \pi} \,
                             \tilde{J}_{\a\g}(\o;\Lambda) \tilde{J}^*_{\a\g}(\o';\Lambda) \\
                           & \times \mr{sinc}\left[\frac{\o-\oD}{2}(t-t_0)\right] \mr{sinc}\left[\frac{\o'-\oD}{2}(t-t_0)\right] \\
                           & \times \mr{sinc}\left[\frac{\o-\oL+E_{\a}-E_{\g}}{2}(t-t_0)\right]
                             \mr{sinc}\left[\frac{\o'-\oL+E_{\a}-E_{\g}}{2}(t-t_0)\right],
    \end{split}
    \label{eq:fermi-gr-sinc}
  \end{equation}
\end{widetext}
where $\mr{sinc}(x) \equiv \sin(x)/x$ denotes the cardinal sine function.

For most practical purposes it is sufficient to pass to the limit of macroscopically long observation times,
corresponding to the limiting case $t-t_0 \gg 2/\oL,2/\oD$.
In this limit, the frequency integrations in Eq.~(\ref{eq:fermi-gr-sinc}) can be simplified considerably
by noting that the two cardinal sine functions in the middle line become highly oscillatory
and hence sharply centered around $\o^{(\prime)} = \oD$.
The $\o$- and $\o'$-integrals can then be well approximated by evaluating
the prefactors in the integrand at this value and pulling them outside the integral:
\begin{widetext}
  \begin{equation}
    \begin{split}
      P_{\mr{inel.}} \xrightarrow[(t-t_0)\gg\frac{2}{\oD}]{}& \ \G_{\mr{kin.}}(\oL,\oD) (t-t_0)^2 \frac{1}{Z_{\mr{M}}} \sum_{\a,\g} \mr{e}^{-\b E_{\g}}
                              \left| \tilde{J}_{\a\g}(\oL - E_{\a} + E_{\g}; \Lambda) \right|^2
                              \mr{sinc}^2\left[\frac{\oL - \oD - E_{\a} + E_{\g}}{2}(t-t_0)\right] \\
                            & \times \left\{ (t-t_0) \int \frac{\d \o}{2 \pi} \, \mr{sinc}\left[\frac{\o-\oL+E_{\a}-E_{\g}}{2}(t-t_0)\right] \right\}^2.
    \end{split}
  \end{equation}
\end{widetext}
The integral in the last line amounts to one and the remaining cardinal sine functions approach
a $\delta$-function in the long-time limit~\cite{cohen1977}:
\begin{equation}
  \mr{sinc}^2(\o t) = \frac{\sin^2(\o t)}{(\o t)^2} \xrightarrow[t \gg \frac{1}{\o}]{} \frac{\pi}{t} \delta(\o).
\end{equation}
Introducing the scattering \emph{rate} via $\dot{P}_{\mr{inel.}} \equiv P_{\mr{inel.}}/(t-t_0)|_{(t-t_0)\oD/2 \to \infty}$,
which is the sensible observable in the long-time limit, we find the considerably simplified result
\begin{widetext}
  \begin{equation}
    \dot{P}_{\mr{inel.}} \xrightarrow[(t-t_0)\gg\frac{2}{\oD}]{} \G_{\mr{kin.}}(\oL,\oD) \frac{1}{Z_{\mr{M}}} \sum_{\a,\g} \mr{e}^{-\b E_{\g}}
                         \left| \tilde{J}_{\a\g}(\oD; \Lambda) \right|^2 \times 2 \pi \delta(\oL - \oD - E_{\a} + E_{\g}).
    \label{eq:fermi-gr}
  \end{equation}
\end{widetext}
This expression has the form of Fermi's golden rule~\cite{cohen1977} with a frequency-dependent matrix element, defined in Eq.~(\ref{eq:me-freq}).
As such, it can be viewed as a generalized Fermi's golden rule beyond first-order time-dependent perturbation theory.
Moreover, the fact that this result still depends on the \emph{complete} matter Hamiltonian and its exact eigenstates
shows that it captures \emph{arbitrary} matter excitations in the final state.
This is in contrast to existing treatments of inelastic light scattering in the literature,
which focus on one matter excitation, most commonly lattice vibrations, at a time.
As such, the above expression generalizes and unifies existing treatments of Raman scattering.

In order to arrive at a practically useful formulation, however, an expression for the matrix elements
$\tilde{J}_{\a\g}(\oD; \Lambda)$ needs to be obtained and the sum over all matter states needs to be evaluated.
In the following we demonstrate how both can be achieved in the low-temperature ($T\to0$) limit by
making use of the Lehmann-Symanzik-Zimmermann (LSZ) reduction formula~\cite{lehmann1955},
which has seen much use in the context of high-energy physics but is not so well-known
in the context of condensed matter physics.


\section{LSZ reduction formula for general matter excitations}
\label{sec:lsz}

In the zero-temperature limit, the sum over $\g$ in Eq.~(\ref{eq:fermi-gr}) reduces to
the term involving the ground state $|0\>$ of $\H_{\mr{M}}$ only.
The only matrix elements needed are then $\tilde{J}_{\a0}(\oD; \Lambda)$, with $|\a\>$ an arbitrary eigenstate of the matter Hamiltonian.
For $|\a\> = |0\>$, this matrix element is simply the exact time-ordered current-current correlation function,
which can be obtained with standard perturbative methods~\cite{abrikosov1965,fetter1971,peskin1995,mahan2000,schwabl2005,weinberg2005}.
For $|\a\> \neq |0\>$, i.e., excited states of the matter system, however,
a different approach is needed and in the following we will show how the LSZ reduction formula can be utilized
to obtain the matrix elements $\tilde{J}_{\a0}(\o; \Lambda)$ for $|\a\> \neq |0\>$.

In an interacting electron-nuclei system, an exact eigenstate $|\a\>$ of the full matter Hamiltonian $\H_{\mr{M}}$ will
involve a mixture of both electronic and vibrational excitations,
including collective excitations such as plasmons, excitons, or phonons.
A non-vanishing matrix element $\<\a| \mathcal{T}\left[ \hat{J}_{\cdots}(t) \hat{J}\dag_{\cdots}(0) \right] |0\>$
then suggests that there exists a correlation between the creation of an electronic or vibrational excitation
and the creation and annihilation of one current each.
As such, we expect on physical grounds that there exists a relation between the needed matrix elements for $|\a\> \neq |0\>$
and a correlation function that involves an electronic or vibrational excitation and two current density operators.
This link can be mathematically established in a precise way by making use of the LSZ reduction formula~\cite{lehmann1955}.
To this end, we define the time-ordered correlation function
\begin{equation}
  S_{O_i}(t',t;\Lambda) \equiv \<0| \mathcal{T}\left[ \hat{O}_i(t') \hat{J}_{\kout,\n}(t) \hat{J}\dag_{\kin,\m}(0) \right] |0\>,
  \label{eq:scat-corfun}
\end{equation}
where $\{\hat{O}_i\}$ denotes a family of (Hermitian) matter operators that describe a matter excitation
and possess non-vanishing matrix elements $\<\a| \hat{O}_i |0\>$
\footnote{We dropped the subscript $I$ for the current operators as with respect to the matter Hamiltonian $\H_{\mr{M}}$ only,
these operators are in the Heisenberg picture.}.
The index $i$ is understood as a generic indexing symbol and can take on both continuous or discrete values,
depending on the nature of the operator $\hat{O}$.
Suitable choices for the $\hat{O}_i$ are the nuclear displacement operators $\uu_{I,i} \equiv \hat{R}_{I,i} - R^{(0)}_{I,i}$,
with $\{\mathbf{R}^{(0)}_{I}\}$ being some equilibrium nuclear configuration, or the electronic charge density operator,
$\hat{\varrho}^{(\mr{el.})}(\r) \equiv (-e)\sum_i \delta^{(3)}(\r-\rr_i)$.
Note that in the \emph{interacting} electron-nuclei system, both of these choices have non-vanishing matrix elements $\<\a|\hat{O}_i|0\>$
and lead to the exact same final result for the Raman scattering rate in case no further approximations are applied.

In a practical treatment, however, one will usually approximate the eigenstates $|\a\>$ of the \emph{fully interacting}
matter Hamiltonian by eigenstates of a non-interacting Hamiltonian describing quasi-particles.
In our, in principal exact, approach, it is then sensible to already choose the $\hat{O}_i$ with future approximations in mind.
For example, if one is primarily interested in matter excitations $|\a\>$ that consist mostly of one-phonon-like excitations,
one would choose $\{\hat{O}_i\} = \{\uu_{I,i}\}$, as then the overlap $\<\a|\hat{O}_{I,i}|0\>$ will be largest.
This will then greatly simplify the later perturbative treatment as discussed below (see also Appendix~\ref{sec:app-oneph}).
Likewise, for states $|\a\>$ involving mostly two-phonon-like excitations,
one would choose $\{\hat{O}_i\} = \{\uu_{I,i}\uu_{J,j}\}$ (cf. Appendix~\ref{sec:app-twoph}).
We want to stress once more that, as long as one always considers the \emph{exact} correlation function $S_{O_i}$,
the LSZ reduction formula will yield the \emph{full} Raman spectrum if $\<\a|\hat{O}_i|0\> \neq 0$
for all eigenstates of the full matter Hamiltonian $\H_{\mr{M}}$.
However, for the purpose of obtaining practically useful computational schemes,
further approximations need to be applied, as discussed in the following, and the $\hat{O}_i$ should be chosen
such that $\<\a|\hat{O}_i|0\>$ is maximized for the states $|\a\>$ one is primarily interested in.

We now return to the task of establishing a precise connection between the correlation function $S_{O_i}$
and the matrix elements $\tilde{J}_{\a 0}(\oD;\Lambda)$.
As pointed out by Lehmann~\textit{et al.}~in a paper about renormalization in quantum electrodynamics~\cite{lehmann1955},
there exists a general connection between scattering matrix elements and residues of correlation functions.
In the context of the present work, we seek to establish a relation between the matrix elements
$\<\a| \mathcal{T}\left[ \hat{J}_{\cdots}(t) \hat{J}\dag_{\cdots}(0) \right] |0\>$ and the residues of the poles
of the Fourier transform of $S_{O_i}$ with respect to its first time argument,
\begin{equation}
  \tilde{S}_{O_i}(\o',t;\Lambda) \equiv \int_{-\infty}^{+\infty} \d t' \, \mr{e}^{i \o' t'} S_{O_i}(t',t;\Lambda).
  \label{eq:s-four}
\end{equation}
Note that $S_{O_i}(t',t;\Lambda)$ is a time-ordered correlation function
and can be evaluated with standard methods from time-dependent perturbation theory
~\cite{abrikosov1965,fetter1971,peskin1995,mahan2000,schwabl2005,weinberg2005}.
Hence the problem of obtaining the matrix elements $\tilde{J}_{\a0}(\o; \Lambda)$
is reduced to finding a suitable expression for $\tilde{S}_{O_i}(\o',t;\Lambda)$
and establishing the precise relation between the residues of its poles to the $\tilde{J}_{\a0}(\o; \Lambda)$.

To establish the latter, we follow Refs.~\cite{peskin1995} and \cite{weinberg2005}
and partition the integral in Eq.~(\ref{eq:s-four}) into three parts by choosing a time $T$
such that $-T < \min \{ t,0 \} \leq \max \{ t,0 \} < +T$ and writing $(-\infty,+\infty) = (-\infty,-T) \cup [-T,+T] \cup (+T,-\infty)$:
\begin{equation}
  \begin{split}
    \tilde{S}_{O_i}(\o',t;\Lambda) =& \int_{-\infty}^{-T} \d t' \, \mr{e}^{i (\o'-i\eta) t'} S_{O_i}(t',t;\Lambda) \\
                                 & + \int_{+T}^{+\infty} \d t' \, \mr{e}^{i (\o'+i\eta) t'} S_{O_i}(t',t;\Lambda) \\
                                 & + \int_{-T}^{+T} \d t' \, \mr{e}^{i \o' t'} S_{O_i}(t',t;\Lambda),
  \end{split}
\end{equation}
where $\eta \equiv 0^+$ is a positive infinitesimal that ensures the convergence of the integrals.
Viewed as a function of $\o'$, the last term does not contribute to the singularity structure of $\tilde{S}_{O_i}$
since the integration is over a finite interval and the integrand is an analytic function of $\o'$.
The first and second terms can be evaluated by using the fact that, in the time-ordered product,
$t'$ is the earliest or latest time, respectively, which ensures that $\hat{O}_i(t')$
always appears as the left- or right-most operator, respectively.
After inserting a complete set of matter states, the Fourier transform reads
\begin{widetext}
  \begin{equation}
    \begin{split}
      \tilde{S}_{O_i}(\o',t;\Lambda) = \sum_ {\a}\bigg\{& i \mr{e}^{i(\o'-\Delta E_{\a}+i\eta)T} \<0|\hat{O}_i|\a\>
                                    \frac{\<\a| \mathcal{T}\left[ \hat{J}_{\kout,\n}(t) \hat{J}\dag_{\kin,\m}(0) \right] |0\>}
                                         {\o' - \Delta E_{\a} +i\eta} \\
                                    & -i \mr{e}^{-i(\o'+\Delta E_{\a}-i\eta)T}
                                      \frac{\<0| \mathcal{T}\left[ \hat{J}_{\kout,\n}(t) \hat{J}\dag_{\kin,\m}(0) \right] |\a\>}
                                           {\o' + \Delta E_{\a} -i\eta} \<\a|\hat{O}_i|0\>
                                      \ \ + \ \ \text{terms regular in $\o'$} \bigg\},
    \end{split}
    \label{eq:s-lehmann}
  \end{equation}
\end{widetext}
where $\Delta E_{\a} \equiv E_{\a} - E_0 > 0$ denotes the exact matter excitation energies.
From Eq.~(\ref{eq:s-lehmann}), it is straightforward to extract the residues of the poles of $\tilde{S}_{O_i}$
for positive frequencies (\emph{Stokes} shift) and to consequently obtain the matrix elements $\tilde{J}_{\a0}(\o; \Lambda)$ as
\begin{equation}
  \begin{split}
    \tilde{J}_{\a0}(\o; \Lambda) =& \lim_{\o'\to\Delta E_{\a}-i\eta} \left\{(\o'-\Delta E_{\a}+i\eta)\tilde{S}_{O_i}(\o',\o;\Lambda)\right\} \\
                               & \times \frac{-i}{\<0|\hat{O}_i|\a\>}.
  \end{split}
  \label{eq:lsz}
\end{equation}
Here, the double Fourier transform of $S_{O_i}$ has been defined in the obvious way as
\begin{equation}
  \tilde{S}_{O_i}(\o',\o;\Lambda) \equiv \int_{-\infty}^{+\infty} \d t \, \mr{e}^{i \o t}
                                      \int_{-\infty}^{+\infty} \d t' \, \mr{e}^{i \o' t'} S_{O_i}(t',t;\Lambda).
\end{equation}
\\
Having obtained an expression for the needed matrix elements in terms of a time-ordered correlation function,
we now turn to the evaluation of the latter.
Note that, given the large nuclei masses, the two current density operators $\JJ$ in $S_{O_i}$
can very well be replaced by the purely electronic density current operators (first term in Eq.~(\ref{eq:current})) only.
The function $S_{O_i}$ then describes the correlation between the creation and annihilation of one electronic current each
and the observable represented by the operator $\hat{O}_i$.
If $\hat{O}_i$ is chosen, for example, as a nuclei displacement operator, the correlation function $S_{O_i}$
will be a mixed electron-nuclei/phonon correlation function.
For a practical application, however, it is useful to work with correlation functions that involve electronic operators only.
This can be achieved by exploiting the fact that for all sensible choices for $\hat{O_i}$,
the Hamiltonian will either involve a direct coupling term of the form
$\sum_j \hat{O}_j \hat{Q}_j$ with $\hat{Q}_j$ being a purely electronic operator
or it will dynamically generate an indirect coupling of $\hat{O}_i$ to an electronic correlation function
(i.e., in higher-order perturbation theory).

The most important example is the case of one-phonon-mediated Raman scattering,
for which one would choose $\{\hat{O}_i\} = \{\uu_{I,i}\}$.
The lowest-order coupling to the electronic sector is then given by the coupling term
\begin{equation}
  \begin{split}
    &\sum_{i,I} \frac{-Z_Ie^2}{|\rr_i-\RR_I|} = \sum_I \int \d^3 r \, \frac{Z_Ie \, \hat{\varrho}^{(\mr{el.})}(\r)}{|\r-\RR_I|} \\
    \supset& \sum_{I} \hat{\mf{u}}_I \cdot \left.\left[ \int \d^3 r \,
    \frac{\pd}{\pd \R_I} \frac{Z_Ie \, \hat{\varrho}^{(\mr{el.})}(\r)}{|\r-\R_I|} \right] \right|_{\Roset},
  \end{split}
  \label{eq:elph}
\end{equation}
which has the form $\sum_j \hat{O}_j \hat{Q}_j$, with $\{\hat{Q}_j\}$
being given by the set of operators $\{\hat{F}_{I,i}\}$ defined as
\begin{equation}
  \hat{F}_{I,i} \equiv \left.\left[ \int \d^3 r \,
  \frac{\pd}{\pd R_{I,i}} \frac{Z_Ie \, \hat{\varrho}^{(\mr{el.})}(\r)}{|\r-\R_I|} \right] \right|_{\Roset}
  \label{eq:elecforce}
\end{equation}
and which represent the electronic force on the nuclei.
Note that for two-phonon-mediated Raman scattering, one would rather choose $\{\hat{O}_i\} = \{\uu_{I,i} \uu_{J,j}\}$
and the coupling to a family of electronic operators $\{\hat{Q}_j\}$ does not fully appear on the level of the Hamiltonian only
(in the form of $\hat{Q}_j \sim \pd^2 \hat{V}_{\mr{e-n}}/ \pd R^{(0)}_{I,i} \pd R^{(0)}_{J,j}$),
but also receives an important contribution from the second-order coupling via the term $\sum_{I,i} \uu_{I,i}\hat{F}_{I,i}$.
A brief discussion of the treatment of two-phonon Raman scattering within the presented formalism can be found in Appendix~\ref{sec:app-twoph}.
In the following discussion though, we will focus on the case where the linear coupling $\sum_j \hat{O}_j \hat{Q}_j$ is sufficient,
as it is the case for one-phonon-mediated Raman scattering.

If we treat the interaction $\sum_j \hat{O}_j \hat{Q}_j$ as the interaction Hamiltonian,
a diagrammatic analysis of the full perturbation series shows that to ``leading order'' in the interaction,
the correlation function $S_{O_i}$ factorizes~\cite{reichardt2018}:
\begin{equation}
  \begin{split}
  S_{O_i}(t',t;\Lambda) \simeq& \sum_j \int_{-\infty}^{+\infty} \d t'' \, \<0| \mathcal{T}\left[ \hat{O}_i(t')  \hat{O}_j(t'') \right] |0\> \\
                      & \times (-i)\<0| \mathcal{T}\left[ \hat{Q}_j(t'') \hat{J}_{\kout,\n}(t) \hat{J}\dag_{\kin,\m}(0) \right] |0\> \\
                     =& i\sum_j \int_{-\infty}^{+\infty} \d t'' \, G_{O_i,O_j}(t',t'') \cM_{Q_j}(t'',t;\Lambda),
  \end{split}
  \label{eq:lo-approx}
\end{equation}
where with
\begin{equation}
  G_{O_i,O_j}(t,t') \equiv (-i) \<0| \mathcal{T}\left[ \hat{O}_i(t)  \hat{O}_j(t') \right] |0\>
\end{equation}
and
\begin{equation}
  \begin{split}
    &\cM_{Q_j}(t',t;\Lambda) \\
    & \qquad \equiv (-i) \left. \<0| \mathcal{T}\left[ \hat{Q}_j(t')
    \hat{J}_{\kout,\n}(t) \hat{J}\dag_{\kin,\m}(0) \right] |0\> \right|_{\mr{connect.}}
  \end{split}
\end{equation}
we defined the \emph{reduced matrix element} $\cM_{Q_j}$ and the exact ``$O$''-excitation Green's function $G_{O_i,O_j}$.
In the example of $\{\hat{O}_i\} = \{\uu_{I,i}\}$, $G_{O_i,O_j}$ is simply equivalent
to the exact phonon propagator in a solid-state physics context,
while $\cM_{Q_j}$, with $\{\hat{Q}_j\} = \{\hat{F}_{I,i}\}$,
is a correlation function that involves purely electronic operators only.
The subscript ``connect.'' refers to the fully \textit{fully connected} part of the correlation function only,
which can mathematically be defined via~\footnote{
We assume ${\<}0|\hat{J}_{\k,\m}|0{\>}=0$, which, for example, is the case in systems with time-reversal or space-inversion symmetry.}
\begin{equation}
  \begin{split}
    & \left. \<0| \mathcal{T}\left[ \hat{Q}_j(t') \hat{J}_{\kout,\n}(t) \hat{J}\dag_{\kin,\m}(0) \right] |0\> \right|_{\mr{connect.}} \\
    &\qquad \equiv \<0| \mathcal{T}\left[ \hat{Q}_j(t') \hat{J}_{\kout,\n}(t) \hat{J}\dag_{\kin,\m}(0) \right] |0\> \\
    & \qquad \phantom{\equiv} - \<0| \hat{Q}_j(t') |0\> \<0| \mathcal{T}\left[ \hat{J}_{\kout,\n}(t) \hat{J}\dag_{\kin,\m}(0) \right] |0\>.
  \end{split}
\end{equation}
For inelastic light scattering, the last, \textit{disconnected} part does not contribute.
Its inclusion is needed, however, to correctly establish links to other approaches to Raman scattering,
such as the method of static, first derivatives of the dielectric susceptibility (cf. Appendix~\ref{sec:app-staticlimit}).
Note that in this approximation, both factors are still \emph{exact} matter correlation functions, i.e.,
the effect of interactions on the two factors are still captured \emph{exactly}.
This is important in order to correctly capture the pole structure of the Fourier transform of $S_{O_i}$,
yet so far has not been given much attention in the existing \textit{ab initio} treatments of Raman scattering.

After Fourier transforming with respect to both $t$ and $t'$,
the needed correlation function becomes a simple product of Fourier transforms,
\begin{equation}
  \tilde{S}_{O_i}(\o',\o;\Lambda) = i \sum_j \tilde{G}_{O_i,O_j}(\o') \tilde{\cM}_{Q_j}(\o',\o;\Lambda),
\end{equation}
with the Fourier transforms of $G_{O_i,O_j}$ and $\cM_{Q_j}$ being defined as
\begin{align}
  \tilde{G}_{O_i,O_j}(\o) &\equiv \int_{-\infty}^{+\infty} \d t \, \mr{e}^{i \o t} G_{O_i,O_j}(t,0) \\
  \tilde{\cM}_{Q_j}(\o',\o;\Lambda) &\equiv
  \begin{aligned}[t]
    \int_{-\infty}^{+\infty} \d t \, \mr{e}^{i \o t} \int_{-\infty}^{+\infty} & \d t' \, \mr{e}^{i \o' t'} \\
    & \times\cM_{Q_j}(t',t;\Lambda). \label{eq:redmat-four}
  \end{aligned}
\end{align}
Making use of the Lehman representation of the exact ``$O$''-excitation Green's function (which is bosonic for all suitable choices of $\hat{O}_i$),
\begin{equation}
  \begin{split}
    \tilde{G}_{O_i,O_j}(\o) = \sum_{\a} \bigg\{ & \frac{\<0|\hat{O}_i|\a\>\<\a|\hat{O}_j|0\>}{\o - \Delta E_{\a} + i\eta} \\
                                & -\frac{\<0|\hat{O}_j|\a\>\<\a|\hat{O}_i|0\>}{\o + \Delta E_{\a} - i\eta}\bigg\},
  \end{split}
  \label{eq:ga-lehmann}
\end{equation}
the matrix element that enters the generalized Fermi's golden rule, Eq.~(\ref{eq:fermi-gr}), reduces to
\begin{equation}
  \tilde{J}_{\a0}(\o; \Lambda) = \sum_j \<\a|\hat{O}_j|0\> \tilde{\cM}_{Q_j}(\Delta E_{\a},\o;\Lambda)
  \label{eq:me-lsz}
\end{equation}
after application of the LSZ reduction formula, Eq.~(\ref{eq:lsz}).
It should once more be pointed out that for the application of the reduction formula,
it is essential to consider interaction effects on the ``$O$''-excitation propagator exactly,
so that the Fourier transform of the latter yields the correct poles for the correlation function $S_{O_i}$.

Finally, we turn back to the generalized Fermi's golden rule for the scattering rate.
Combining Eqs.~(\ref{eq:fermi-gr}) and (\ref{eq:me-lsz}),
we obtain the following compact, yet insightful expression for the rate for Raman, i.e., inelastic light scattering:
\begin{equation}
  \begin{split}
    \dot{P}_{\mr{inel.}} \simeq & \ \G_{\mr{kin.}}(\oL,\oD) \sum_{i,j} \tilde{\cM}^*_{Q_i}(\oL-\oD,\oD;\Lambda) \\
                                & \times 2\pi \cA_{O_i,O_j}(\oL-\oD) \tilde{\cM}_{Q_j}(\oL-\oD,\oD;\Lambda).
  \end{split}
  \label{eq:rate-spectralfun}
\end{equation}
Here we identified the ``$O$''-excitation spectral function as
\begin{equation}
  \cA_{O_i,O_j}(\o) = \sum_{\a} \<0|\hat{O}_i|\a\> \<\a|\hat{O}_j|0\> \delta(\o-\Delta E_{\a}),
\end{equation}
assuming $\o = \oL - \oD > 0$, since in the $T\to0$-limit, only matter excitations in the final state contribute, i.e.,
the frequency of the scattered and detected light is \emph{decreased}
with respect to the frequency of the incoming photon (\emph{Stokes} shift).
Note that for $\o>0$, the spectral function and time-ordered Green's function are related by the identity
$\cA_{O_i,O_j}(\o) = -\frac{1}{\pi}\mr{Im}\,\tilde{G}_{O_i,O_j}(\o)$, as follows directly from Eq.~(\ref{eq:ga-lehmann}).


\section{Discussion and outlook}
\label{sec:discussion}

Besides the presented correlation function-based approach, Eq.~(\ref{eq:corfun}),
the derivation of Eq.~(\ref{eq:rate-spectralfun}) and the demonstration of the usefulness of the LSZ reduction formula
in a condensed matter physics context and beyond its high-energy physics origins constitute the major results presented in this work.
The derived expression for the Raman scattering rate given in Eq.~(\ref{eq:rate-spectralfun})
has many theoretical, conceptional, and practical advantages over the approaches used in the works mentioned in the introduction.

From a theoretical and conceptual point of view, our expression allows a decoupling of the line shape and the intensity/area of the Raman peaks.
They can thus be studied and modeled independently and in a theoretically consistent and well-defined way.
While the appearance of the spectral function ensures that possible satellite peaks are captured in the description as well,
the reduced matrix element ensures that the electronic response is captured exactly.
The latter in particular includes both possible non-adiabatic effects due to the two-frequency nature of the reduced matrix element
and non-adiabatic effects due to being based on an \emph{exact} correlation function.

From a practical perspective, this expression can be used to devise a clear computational recipe
for the development of a future \textit{ab initio} code for the computation of
Raman intensities including both excitonic and non-adiabatic effects~\cite{reichardt2018c}.
This is of great interest in condensed matter physics.
There the most prominent features in the Raman spectrum are due to the excitation of phonons,
which can usually be treated as \emph{quasi-particles (QPs)} with a finite frequency $\o_{\l}$ and a finite decay width $\Gamma_{\l}$.
For $\o=\oL-\oD$ near $\o_{\l}$, the spectral function can be approximated as a Lorentzian
\begin{equation}
  \cA_{O_i,O_j}(\o) \stackrel{\o \to \o_{\l}}{\simeq}
    v_i^{\l} \frac{Z_{\l}}{\pi}\frac{\G_{\l}/2}{(\o-\o_{\l})^2 + (\G_{\l}/2)^2} \Big(v_j^{\l}\Big)^*.
\end{equation}
Here, $Z_{\l}$$\leq$$1$ is the QP-weight, which accounts for the fact that a ``phonon'' excitation
in an interacting electron-nuclei system is actually a mixture of correlated nuclear displacements and electronic excitations,
while $v_i^{\l}$ is the eigenvector associated with the QP labeled by $\l$.
In this quasi-particle-approximation (QPA) then, the Raman spectrum reduces to a weighted sum over Lorentzian peaks:
\begin{equation}
  \begin{split}
    \dot{P}_{\mr{inel.}} \stackrel{\text{QPA}}{\approx} & \G_{\mr{kin.}}(\oL,\oD) \sum_{\l} \left| \tilde{\cM}^{(\mr{QP})}_{\l}(\oL-\oD,\oD;\Lambda) \right|^2 \\
                                                        & \times \frac{Z_{\l}}{\pi}\frac{\G_{\l}/2}{(\oL-\oD-\o_{\l})^2 + (\G_{\l}/2)^2},
  \end{split}
\end{equation}
where
\begin{equation}
  \tilde{\cM}^{(\mr{QP})}_{\l}(\o',\o;\Lambda) \equiv \sum_j \Big( v^{\l}_j \Big)^* \tilde{\cM}_{Q_j}(\o',\o;\Lambda)
  \label{eq:redmat-proj}
\end{equation}
is the reduced matrix element projected onto the QP-eigenvector.
Note that all quantities can be computed within many-body perturbation theory since they are all derived from time-ordered correlation functions.
For instance, the QP-frequency and -decay widths as well as the QP-weights can be obtained
from the corresponding self-energy and Dyson's equation for the Green's function,
while the reduced matrix element can be computed with perturbative and/or diagrammatic methods within the desired approximation.

For a first demonstration of the actual usefulness of the derived expressions in a computational sense,
we demonstrate in Appendix~\ref{sec:app-oneph} how the outlined approach reduces to the previously used
\textit{ab initio}, independent-particle theory of one-phonon Raman scattering~\cite{reichardt2017b} in the corresponding approximations.
The practical inclusion of excitonic effects, which is already built-in in the presented theoretical formalism
will be the subject of a further work, which is already in preparation~\cite{reichardt2018c}.
For completeness, we briefly sketch the treatment of ``second-order'', i.e., two-phonon-mediated,
Raman scattering on the independent-particle level in Appendix~\ref{sec:app-twoph}.

The outlined approach for the calculation of Raman intensities has its biggest potential for condensed matter systems,
where it allows one to capture both non-adiabatic and excitonic effects at the same time.
The combination of both effects can be expected to be important especially in low-dimensional, light, small-band gap materials.
Typically examples are carbon nanotubes, in which the band gap can be on the same order of magnitude
as the phonon frequencies while excitonic effects play a non-negligible role~\cite{varsano2017}.
In addition, the suggested approach has the potential to be of great use for the automated screening of materials
with yet unknown optical and vibrational properties in high-throughput searches, for which it is \textit{a priori}
not known if excitonic and non-adiabatic effects are negligible or not.
Computationally, our method allows all calculations to be done within the unit cell of the equilibrium structure
and without breaking any symmetries as is done in static finite-difference approaches.
The latter not only break symmetries when displacing atoms, thus increasing the computational time, but also require convergence studies
with respect to the smallness of the displacement used to emulate the derivative with respect to atomic displacements.
As such, our work also proves to be a big first step towards a more comprehensive, yet also computationally more efficient approach
that is also highly modular in its theoretical and algorithmic structure.


\section{Conclusions}
\label{sec:concl}

In the present work, we have presented a generalized and fully quantum mechanical treatment of inelastic light scattering.
While the correlation-function approach has the advantage of being completely general and being valid also on ultra-short time scales,
we also derived a generalized version of Fermi's golden rule to obtain a practically useful expression.
In addition, we further showed how the LSZ reduction formula can usefully be applied
outside its usual high-energy physics context to obtain the matrix elements in the generalized Fermi's golden rule
from time-ordered correlation functions, which constitute the main backbone
of modern condensed matter physics theory and \textit{ab initio} descriptions.
This formulation allowed us to obtain an expression for the Raman scattering rate
entirely in terms of time-ordered correlation and spectral functions.
It thus permits a systematic inclusion of many-body effects,
such as excitonic intermediate states and non-adiabatic electron-phonon interactions~\cite{reichardt2018},
and also paves the way toward a computationally feasible \textit{ab initio} implementation~\cite{reichardt2018c}
beyond the current state of the art.


\begin{acknowledgments}

The authors would like to thank A.~Marini for initial inspiring discussions as well as
M.~Sadhukhan and Y.~Al-Hamdani for helpful discussions during the revision of the manuscript.
S.R. and L.W. acknowledge financial support by the National Research Fund (FNR) Luxembourg
(projects RAMGRASEA and INTER/ANR/13/20/NANOTMD).
S.R. also acknowledges financial support by the Leverhulme Trust (Grant RL-2012-001).

\end{acknowledgments}


\appendix

\section{Connection to the non-adiabatic theory of one-phonon-induced
         Raman scattering on the independent-particle level}
\label{sec:app-oneph}

As an illustration and to provide a first example of the computational feasibility of the presented method,
we show in this Appendix, how concrete formulas for the previously used non-adiabatic, independent-particle theory for one-phonon-induced Raman scattering
can be obtained from the presented theoretical approach.
The more comprehensive version and implementation of the presented approach that also captures excitonic effects is subject of future work.

To obtain concrete expressions for the case of one-phonon-mediated Raman scattering
with electrons being treated on the level of independent particles, we start from Eq.~(\ref{eq:rate-spectralfun}).
For the family of operators $\{\hat{O}_i\}$ we choose the nuclei displacement operators $\{\uu_{I,i}\}$
and approximate the needed exact correlation function $S_{O_i}$, Eq.~(\ref{eq:scat-corfun}),
in the ``leading-order'' coupling approximation, Eq.~(\ref{eq:lo-approx}).
As given in Eq. (\ref{eq:elph}), the role of the operators $\{\hat{Q}_i\}$
is then taken on by the electronic force operators $\{\hat{F}_{I,i}\}$, defined in Eq.~(\ref{eq:elecforce}).
The needed spectral function and reduced matrix element for use in Eq.~(\ref{eq:rate-spectralfun}) are then
the phonon spectral function $\cA_{u_{I,i},u_{J,j}}(\o)$ and the time-ordered correlation function describing the
correlation between an electronic force on the nuclei and creation and annihilation of one electronic current each:
\begin{equation}
  \cM_{F_{I,i}}(t',t;\Lambda) = (-i) \<0| \mathcal{T}\left[ \hat{F}_{I,i}(t') \hat{J}_{\kout,\n}(t) \hat{J}\dag_{\kin,\m}(0) \right] |0\>.
  \label{eq:redmat_elec}
\end{equation}

As mentioned in Section~\ref{sec:discussion}, the spectral function for $\omega$
near a one-phonon excitation energy $\o_{\q,\l}$ can typically be rather accurately approximated
by the quasi-particle phonon spectral function
\begin{equation}
  \begin{split}
    \cA^{(\mr{QP})}_{u_{I,j},u_{J,j}}&(\o) \stackrel{\o \to \o_{\q,\l}}{\simeq} \\
    &v_{I,i}^{\q,\l} \frac{Z_{\q,\l}}{\pi}\frac{\G_{\q,\l}/2}{(\o-\o_{\q,\l})^2 + (\G_{\q,\l}/2)^2} \Big(v_{J,j}^{\q,\l}\Big)^*.
  \end{split}
\end{equation}
In a crystalline solid, the phonon eigenvectors read
\begin{equation}
  v_{I,i}^{\q,\l} = \sqrt{\frac{1}{2 N M_{\a} \o_{\q,\l}}} \mr{e}^{i\q\cdot\R_n} v_{\a,i}^{\q,\l},
\end{equation}
with $\q$ and $\l$ labeling the phonon wave vector and branch, respectively,
$N$ being the number of unit cells of the solid (which is assumed to obey periodic boundary conditions),
$\R_n$ denoting the origin of the unit cell in which atom $I$ is located in the equilibrium structure,
and $\a$ labeling the different atoms within the unit cell.
$v_{\a,i}^{\q,\l}$ is the orthonormalized eigenvector of the dynamical matrix to the eigenvalue $\o^2_{\q,\l}$.
Note that for systems in which the electron-phonon coupling is weak,
the quasi-particle factor $Z_{\q,\l}$ can be well set to unity.

The reduced matrix element is typically approximated by neglecting the nuclei contribution
to the electronic current, defined in Eq.~(\ref{eq:current}).
The electron charge and current density operators can then conveniently be expressed in terms
of the electronic field operators only as
\begin{align}
  \hat{\varrho}^{(\mr{el.})}(\r) &= (-e) \sum_i \delta^{(3)}(\r-\rr_i) \nonumber \\
                                 &= (-e) \ppsi\dag(\r) \ppsi(\r),
                                    \label{eq:el-denop} \\
  \JJ(\r) \approx \JJ^{(\mr{el.})}(\r) &= (-e) \sum_i \delta^{(3)}(\r-\rr_i)\frac{\pp_i}{m} \nonumber \\
                                       &= \frac{(-e)}{m} \ppsi\dag(\r) (-i\boldsymbol{\nabla}) \ppsi(\r).
                                          \label{eq:el-curop}
\end{align}
Combining Eqs.~(\ref{eq:currentfour}), (\ref{eq:elecforce}), (\ref{eq:redmat_elec}), (\ref{eq:el-denop}), and (\ref{eq:el-curop})
then leads to the following expression for the fully Fourier-transformed (Eq.~(\ref{eq:redmat-four}))
and projected (Eq.~(\ref{eq:redmat-proj})) reduced matrix element:
\begin{widetext}
\begin{equation}
  \begin{split}
     & \phantom{\times} \tilde{\cM}^{(\mr{QP})}_{\q,\l}(\o';\o;\Lambda) \\
    =& \phantom{\times} \sum_{n} \frac{1}{\sqrt{N}} \mr{e}^{-i\q\cdot\R_n}
       \int \d^3 r_1 \, \mr{e}^{-i \kout \cdot \r_1}
       \int \d^3 r_2 \, \mr{e}^{+i \kin \cdot \r_2}
       \sum_{\a,k} \sqrt{\frac{1}{2 M_{\a} \o_{\q,\l}}} v_{\a,k}^{\q,\l;*}
       \sum_i \left(\epsilon^{i}_{\kout,\n}\right)^*
       \sum_j \epsilon^j_{\kin,\m} \\
     & \times \lim_{\r'_1 \to \r_1} \frac{-e}{m} (-i) \frac{\pd}{\pd r'_{1,i}}
              \lim_{\r'_2 \to \r_2} \frac{-e}{m} (-i) \frac{\pd}{\pd r'_{2,j}}
              \int \d^3 r_3 \, \left.\left[ \frac{\pd}{\pd R_{n,\a,k}} \frac{-Z_{\a}e^2}{|\r_3-\R_{n,\a}|} \right] \right|_{\Roset}
              \int_{-\infty}^{+\infty} \d t \, \mr{e}^{i \o t}
              \int_{-\infty}^{+\infty} \d t' \, \mr{e}^{i \o' t'} \\
     & \times (-i) \<0| \mathcal{T}\left[ \ppsi\dag(\r_3,t'^+)\ppsi(\r_3,t') \ppsi\dag(\r_1,t^+)\ppsi(\r'_1,t)
                                          \ppsi\dag(\r_2,0^+)\ppsi(\r'_2,0) \right] |0\>.
  \end{split}
\end{equation}
\end{widetext}
In this expression, the first line contains the spatial Fourier transforms and projections on the polarization vectors of the
incoming and outgoing light, the second line contains the gradient operators from the electron-light coupling,
the (bare) electron-phonon coupling, and the temporal Fourier transforms, while the last line consists of
the exact three-electron Green's function of the interacting electron-nuclei system.
Note that the latter contains both excitonic effects and the screening of the electron-phonon coupling~\cite{reichardt2018}.
In the previously used \textit{ab initio} treatments within the independent-particle approximation,
the three-particle Green's function, more precisely its fully connected part, was approximated on the independent-particle level as
\begin{widetext}
\begin{equation}
  \begin{split}
    & \<0| \mathcal{T}\left[ \ppsi\dag(\r_3,t'^+)\ppsi(\r_3,t') \ppsi\dag(\r_1,t^+)\ppsi(\r'_1,t)
         \ppsi\dag(\r_2,0^+)\ppsi(\r'_2,0) \right] |0\> \\
    \simeq& \phantom{ + }(-1) \<0| \mathcal{T}\left[ \ppsi(\r'_1,t)\ppsi\dag(\r_3,t'^+) \right] |0\>
                             \<0| \mathcal{T}\left[ \ppsi(\r'_2,0)\ppsi\dag(\r_1,t^+) \right] |0\>
                             \<0| \mathcal{T}\left[ \ppsi(\r_3,t')\ppsi\dag(\r_2,0^+) \right] |0\> \\
          & + (-1) \<0| \mathcal{T}\left[ \ppsi(\r'_2,0)\ppsi\dag(\r_3,t'^+) \right] |0\>
                   \<0| \mathcal{T}\left[ \ppsi(\r'_1,t)\ppsi\dag(\r_2,0^+) \right] |0\>
                   \<0| \mathcal{T}\left[ \ppsi(\r_3,t')\ppsi\dag(\r_1,t^+) \right] |0\>,
  \end{split}
\end{equation}
\end{widetext}
where the overall minus sign arises from the anti-commuting nature of the field operators.
Note that in this approximation, the screening of the electron-phonon coupling is entirely neglected
and in the previous works this was remedied by replacing the bare electron-phonon coupling with a screened version.
It is possible to show that this procedure is equivalent to approximating the exact three-particle Green's function
by a certain \emph{subseries} of Feynman diagrams instead, which can be summed up to yield a product of the screened electron-phonon coupling
and the independent-particle three-electron Green's function~\cite{reichardt2018}.

The most straightforward and reasonably accurate way of turning this expression into
one that can be computed with modern \textit{ab initio} methods is to further approximate the exact one-electron Green's functions
on the level of Kohn-Sham density functional theory by replacing $|0\> \to |0_{\mr{KS}}\>$ and to
replace the bare electron-phonon interaction by its screened counterpart obtained
using density functional perturbation theory:
\begin{equation}
  \begin{split}
    &\left.\left[ \frac{\pd}{\pd R_{n,\a,k}} \frac{-Z_{\a}e^2}{|\r-\R_{n,\a}|} \right] \right|_{\Roset} \\
    &\ \ \ \to \left.\left[ \frac{\pd}{\pd R_{n,\a,k}} V_{\mr{scf}}(\r;\Rset) \right] \right|_{\Roset},
  \end{split}
\end{equation}
where $V_{\mr{scf}}$ denotes the total self-consistent potential in Kohn-Sham density functional theory.

Finally, concrete expressions that only involve quantities obtainable from computational, \textit{ab initio} calculations
can be obtained by expanding the electron field operator in terms of Kohn-Sham wave functions and annihilation operators:
$\ppsi(\r) = \sum_{\k,n} \phi_{\k,n}(\r) \cc_{\k,n}$, where $\k$ and $n$ denote the wave vector
and band index of a Kohn-Sham one-electron state.
Employing the dipole approximation $\kin=\kout=\mf{0}$ further simplifies the expressions as momentum conversation
enforces $\q=\mf{0}$ as well and all matrix elements between Kohn-Sham states become diagonal in $\k$-space.
We then define the (bare) electron-photon and the (screened) electron-phonon coupling matrix elements as
\begin{widetext}
\begin{align}
  d^{\m}_{\k;m,n} &\equiv \frac{-e}{m} \boldsymbol{\epsilon}_{\k=\mf{0},\m} \cdot
                          \int \d^3 r \, \phi^*_{\k,m}(\r) (-i\boldsymbol{\nabla}) \phi_{\k,n}(\r) \\
  g^{\l}_{\k;m,n} &\equiv \sum_{n,\a,k} \sqrt{\frac{1}{2 N M_{\a} \o_{\q,\l}}} v_{\a,k}^{\q=\mf{0},\l}
  \int \d^3 r \, \phi^*_{\k,m}(\r) \left.\left[ \frac{\pd}{\pd R_{n,\a,k}} V_{\mr{scf}}(\r;\Rset)
                                            \right] \right|_{\Roset} \phi_{\k,n}(\r)
\end{align}
\end{widetext}
and the Fourier-transform of the one-electron Kohn-Sham-level Green's function as
\begin{equation}
  \begin{split}
       \tilde{G}_{\k,a}(\o) &\equiv \int \d t \, \mr{e}^{i \o t} (-i)
       \<0_{\mr{KS}}| \mathcal{T} \left[ \cc_{\k,a}(t) \cc\dag_{\k,a}(0) \right] |0_{\mr{KS}}\> \\
    &= \frac{f_{\k,a}}{\o - \e_{\k,a} - i\g_{\k,a}} + \frac{1-f_{\k,a}}{\o - \e_{\k,a} + i\g_{\k,a}},
  \end{split}
\end{equation}
with $\e_{\k,a}$ denoting the energy, $\g_{\k,a}$ the (positive) decay width,
and $f_{\k,a}$ the ground-state occupation of a Kohn-Sham state.
Using a matrix notation in the space of Kohn-Sham bands, defined in the obvious way,
the expression for the reduced matrix element, including a factor of 2 to account for the electron spin, reads:
\begin{widetext}
\begin{equation}
  \begin{split}
    \tilde{\cM}^{(\mr{QP})}_{\q=\mf{0},\l}(\o';\o;\m,\n) = 2 \sum_{\k} \int \frac{\d \o''}{2\pi} \, \Bigg\{ &
    \mr{tr} \bigg[ \ms{G}_{\k}(\o''+\o'+\o) \ms{d}^{\m}_{\k} \ms{G}_{\k}(\o'')
                   \ms{g}^{\l;\dagger}_{\k} \ms{G}_{\k}(\o''+\o') \ms{d}^{\n;\dagger}_{\k} \bigg] \\
    & + \mr{tr} \bigg[ \ms{G}_{\k}(\o''+\o'+\o) \ms{d}^{\m}_{\k} \ms{G}_{\k}(\o'')
                   \ms{d}^{\n;\dagger}_{\k} \ms{G}_{\k}(\o''+\o) \ms{g}^{\l;\dagger}_{\k} \bigg] \Bigg\},
  \end{split}
\end{equation}
\end{widetext}
which coincides with the expressions previously used in numerical \textit{ab initio}
calculations on the independent-particle level~\cite{reichardt2017b}.
Note, however, that here, $\o$ corresponds to the frequency of the \emph{outgoing} photon,
whereas in Ref.~\onlinecite{reichardt2017b}, $\o$ corresponds to the frequency of the \emph{incoming} photon.


\section{Connection of the presented approach to the static displacement method
\label{sec:app-staticlimit}}

In Appendix~\ref{sec:app-oneph}, we discussed the independent-particle, but non-adiabatic limit.
Here, we show in which limit the full theory reduces to the approach of calculating static first derivatives of the dielectric susceptibility.
This approach captures electronic correlations beyond the independent-particle level, but cannot capture non-adiabatic effects.

The starting point of this formulation is the \textit{adiabatic approximation} to the full matter Hamiltonian of Eq.~(\ref{eq:hmatter}).
Starting from a fixed nuclei configuration $\Rset$, one defines a mean-field Hamiltonian in the adiabatic, Born-Oppenheimer approximation as
\begin{equation}
  \H^{\ad}(\Rset) \equiv \he(\Rset) + \hat{h}_{\mr{n}} - \Eeo(\Rset) \hat{\mathbbm{1}},
\end{equation}
where
\begin{equation}
  \he(\Rset) \equiv \sum_i \frac{\pp^2_i}{2 m} + \frac{1}{2}\sum_{\substack{i,j \\i \neq j}} \frac{e^2}{|\rr_i-\rr_j|}
                    + \sum_{i,I} \frac{-Z_Ie^2}{|\rr_i-\R_I|}
\end{equation}
is a purely electronic Hamiltonian with the electrons moving in a potential provided by static nuclei and
\begin{equation}
  \hat{h}_{\mr{n}} \equiv \sum_I \frac{\PP^2_I}{2 M_I} + \frac{1}{2}\sum_{\substack{I,J \\I \neq J}} \frac{Z_I Z_J e^2}{|\RR_I-\RR_J|}
                          + \Eeo(\RRset),
\end{equation}
is a purely nuclear Hamiltonian with the nulcei moving in the mean-field potential $\Eeo(\Rset)$
defined as the energy of the ground state $|\0\>_{\mr{e}}$ of the electronic Hamiltonian:
\begin{equation}
  \he(\Rset) |\0(\Rset)\>_{\mr{e}} = \Eeo(\Rset) |\0(\Rset)\>_{\mr{e}}.
\end{equation}
We will denote the ground state of the full adiabatic Hamiltonian
$\H^{\ad}(\Rset)$ by $|\0(\Rset)\>=|\0(\Rset)\>_{\mr{e}} \otimes |\0\>_{\mr{n}}$.

In this previous theory of Raman scattering, the expression for the Raman scattering rate reads~\cite{yu2010,gillet2013,miranda2017}:
\begin{equation}
  \begin{split}
    \dot{P}^{\ad}_{\mr{inel.}} = \G_{\mr{kin.}}(\oL,\oD) \sum_{\l}
      & \left| \left. \frac{\pd \tilde{\chi}^{\ad}_{ij}(\oL)}{\pd Q_{\l}} \right|_{\Roset} \right|^2 \\
      & \times 2 \pi \delta(\oL - \oD - \o^{\ad}_{\l}),
  \end{split}
\end{equation}
where $\G_{\mr{kin.}}(\oL,\oD)$ was defined in Eq.~(\ref{eq:gam-kin}).
Here,
\begin{equation}
  \frac{\pd}{\pd Q_{\l}} \equiv \sum_I \mf{v}^{\q=\mf{0},\l}_I \cdot \frac{\pd}{\pd \R_I}
\end{equation}
is the directional derivative along a $\q$=$\mf{0}$ phonon eigenvector $\mf{v}^{\q=\mf{0},\l}_I$~\footnote{
Note that at $\q$=$\mf{0}$, the phonon eigenmodes are entirely fixed by the lattice symmetry
and thus are the same in both the adiabatic and non-adiabatic theory.
We thus do not need to distinguish between the exact $\mf{v}^{\q=\mf{0},\l}_I$
and the adiabatic $\mf{v}^{\ad,\q=\mf{0},\l}_I$.}
and $\tilde{\chi}^{\ad}_{ij}(\o)$ are the cartesian components of the
Fourier-transformed transverse dielectric susceptibility tensor in the dipole approximation:
\begin{equation}
  \tilde{\chi}^{\ad}_{ij}(\o) \equiv \int_{-\infty}^{+\infty} \d t \, \mr{e}^{i \o t}
                                     (-i) \<\0| \mathcal{T} \left[ \hat{J}^{\ad}_i(t) \hat{J}^{\ad}_j(0) \right] |\0\>.
\end{equation}
In the equation above, we defined the Heisenberg picture with respect to the adiabatic Hamiltonian $\H^{\ad}=\H^{\ad}(\Rset)$, i.e.,
\begin{equation}
  \hat{J}^{\ad}_i(t) = \hat{J}^{\ad}_i(t;\Rset) \equiv \mr{e}^{i \H^{\ad}t} \hat{J}_i \mr{e}^{-i \H^{\ad}t},
\end{equation}
with the cartesian components of the electronic current density operator (Eq.~(\ref{eq:el-curop})) in the dipole approximation given by
$\hat{J}_i \equiv \int \d^3 r \, \hat{J}^{(\mr{el}.)}_i(\r)$.
Note that the current-current correlation function only involves electronic operators.
Since the adiabatic Hamiltonian is the sum of a purely electronic and nuclear part,
the nuclear part of the Hamiltonian and the ground state drop out, and we can simply write
\begin{equation}
  \tilde{\chi}^{\ad}_{ij}(\o) = \int_{-\infty}^{+\infty} \d t \, \mr{e}^{i \o t}
                                     (-i) \ {}_{\mr{e}}\<\0| \mathcal{T} \left[ \hat{J}_{i,I}(t) \hat{J}_{j,I}(0) \right] |\0\>_{\mr{e}},
\end{equation}
with the interaction picture operators understood to be defined as
\begin{equation}
  \hat{J}_{i,I}(t) = \hat{J}_{i,I}(t;\Rset) \equiv \mr{e}^{i \he(\Rset) t} \hat{J}_i \mr{e}^{-i \he(\Rset) t}.
\end{equation}

By comparison, our fully quantum mechanical theory in the dipole approximation yields the expression
(compare Section~\ref{sec:lsz} and Appendix~\ref{sec:app-oneph}):
\begin{equation}
  \begin{split}
    &\dot{P}^{\ad}_{\mr{inel.}} = \\
    & \qquad \G_{\mr{kin.}}(\oL,\oD) \sum_{\l} \sum_{\substack{I,J \\ a,b}}
      \left[ \tilde{\cM}_{F_{I,a}}(\oL-\oD,\oL;i,j) \right]^* \\
    & \qquad \times 2 \pi \cA_{u_{I,a},u_{J,b}}(\oL-\oD) \tilde{\cM}_{F_{J,b}}(\oL-\oD,\oL;i,j),
  \end{split}
\end{equation}
where the fully Fourier-transformed reduced matrix element in the dipole approximation reads
\begin{equation}
  \begin{split}
    \tilde{\cM}_{F_{I,a}}(\o',\o;i,j) =& \int_{-\infty}^{+\infty} \d t \, \mr{e}^{i \o t} \int_{-\infty}^{+\infty} \d t' \, \mr{e}^{i \o' t'} \\
      & \times (-i) \<0| \mathcal{T}\left[ \hat{F}_{I,a}(t') \hat{J}_i(t) \hat{J}_j(0) \right] |0\>.
  \end{split}
\end{equation}
Here, the operators are in the Heisenberg-picture with respect to the full matter Hamiltonian (Eq.~(\ref{eq:hmatter})),
whose ground state is denoted by $|0\>$, and we also treat the (electronic) current operators in the dipole approximation.
The force operator $\hat{F}_{I,a}$ has been defined previously in Eq.~(\ref{eq:elecforce}).

We shall now show that our theory reduces to the adiabatic theory of Raman scattering in the limits
$\H_{\mr{M}} \to \H^{\ad}(\Roset)$ and $|0\>\to|\0(\Roset)\>$, i.e.,
we replace the full Hamiltonian and exact ground state by their adiabatic counterparts defined with respect to
the equilibrium nuclei configuration $\Roset$~\footnote{
The proof does not make use of the fact that $\Roset$ represents the nuclei configuration of minimal energy
and also holds for any other configuration $\Rset$, provided the same configuration is used both in
the bare force operators $\hat{F}_{I,a}$ and the adiabatic Hamiltonian and ground state.}.
\emph{However, we will see that the adiabatic theory is only obtained if we additionally
neglect the frequency dependence of the reduced matrix element on $\o'$,
that is, we additionally let $\o' \to 0$ in $\tilde{\cM}_{F_{I,a}}(\o',\o;i,j)$.}
This illustrates once more that the fully dynamic and quantum mechanical theory
contains additional dynamical features and structures in the matrix element that are entirely lost in static approaches.

We will focus our discussion on establishing a link between the reduced matrix element
and the derivative of the dielectric susceptibility along an adiabatic phonon eigenvector.
The phonon spectral function, by contrast, is trivially shown to reduce to the form
\begin{equation}
  \cA_{u_{I,a},u_{J,b}}(\o) \to \sum_{\l} v^{\q=\mf{0},\l}_{I,a} \left(v^{\q=\mf{0},\l}_{J,b}\right)^* \delta(\o-\o^{\ad}_{\l}),
\end{equation}
if only $\q=\mf{0}$-modes are considered, which, as mentioned in Appendix~\ref{sec:app-oneph},
are the only ones that contribute in the dipole approximation.
Note that, strictly speaking, an exact $\delta$-function-like shape of the spectral function
is only obtained by additionally going to the \textit{harmonic approximation} as well,
as otherwise phonon-phonon scattering, i.e., anharmonic effects, will modify this shape.
However, for the purpose of establishing a link between the theory presented here and the adiabatic case,
this is of no concern.
We will then demonstrate below that
\begin{widetext}
\begin{equation}
    \left| \sum_{I,a} \left(v^{\q=\mf{0},\l}_{I,a}\right)^*
    \left. \left[ \tilde{\cM}_{F_{I,a}}(\o'=0,\o;i,j) \right]\right|_{\mr{ad.}} \right|
  = \left| \left. \frac{\pd \tilde{\chi}^{\ad}_{ij}(\o)}{\pd Q_{\l}} \right|_{\Roset} \right|,
  \label{eq:ad-limit-conjecture}
\end{equation}
\end{widetext}
which demonstrates that in the adiabatic limit and when neglecting the frequency dependence of the reduced matrix element,
the full theory reduces to the previously employed adiabatic approach based on static first derivatives.

To start with, it is convenient to define the projected force operator
\begin{equation}
  \hat{F}_{\l} \equiv \sum_{I,a} \left(v^{\q=\mf{0},\l}_{I,a}\right)^* \hat{F}_{I,a}
  = \left. \frac{\pd \he(\Rset)}{\pd Q_{\l}} \right|_{\Roset},
\end{equation}
where the last identity follows from the definition of the projected derivative and the force operator $\hat{F}_{I,a}$.
We then have:
\begin{equation}
  \begin{split}
    &\sum_{I,a} \left(v^{\q=\mf{0},\l}_{I,a}\right)^*
     \left. \left[ \tilde{\cM}_{F_{I,a}}(\o'=0,\o;i,j) \right] \right|_{\mr{ad.}} \\
   =& \int_{-\infty}^{+\infty} \d t \, \mr{e}^{i \o t} \int_{-\infty}^{+\infty} \d t' \, (-i) \\
   & \times \bigg( \<\0| \mathcal{T}\left[ \hat{F}^{\ad}_{\l}(t') \hat{J}^{\ad}_i(t) \hat{J}^{\ad}_j(0) \right] |\0\> \\
   & \ - \left. \<\0| \hat{F}^{\ad}_{\l}(t') |\0\>
     \<\0| \mathcal{T}\left[ \hat{J}^{\ad}_i(t) \hat{J}^{\ad}_j(0) \right] |\0\> \bigg) \right|_{\Roset}.
 \end{split}
 \label{eq:ad-corfun-connect}
\end{equation}
Note that, again, all operators are purely electronic operators and hence all correlation functions
reduce to their purely electronic versions, i.e., we can let $\H^{\ad} \to \he$ and $|\0\>\to|\0\>_{\mr{e}}$.

{
Next, we directly evaluate the $t'$-integral of the first term, which involves the full $F$-$J$-$J$ correlation function.
From the definition of the time-ordered product, we directly have
\begin{widetext}
\begin{equation}
  \begin{split}
     & \int_{-\infty}^{+\infty} \d t' \, \left. \left\{ {}_{\mr{e}}\<\0| \mathcal{T}\left[ \hat{F}_{\l,I}(t')
       \hat{J}_{i,I}(t) \hat{J}_{j,I}(0) \right] |\0\>_{\mr{e}} \right\} \right|_{\Roset} \\
    =& \bigg( \theta(t) \left[   \int_t^{\infty} \d t' \, {}_{\mr{e}}\<\0| \hat{F}_{\l,I}(t') \hat{J}_{i,I}(t) \hat{J}_j |\0\>_{\mr{e}}
                        + \int_0^t \d t' \, {}_{\mr{e}}\<\0| \hat{J}_{i,I}(t) \hat{F}_{\l,I}(t') \hat{J}_j |\0\>_{\mr{e}}
                        + \int_{-\infty}^0 \d t' \, {}_{\mr{e}}\<\0| \hat{J}_{i,I}(t) \hat{J}_j \hat{F}_{\l,I}(t') |\0\>_{\mr{e}} \right] \\
    & \! \! + \theta(-t) \left. \left[   \int_0^{\infty} \d t' \,  {}_{\mr{e}}\<\0| \hat{F}_{\l,I}(t') \hat{J}_j \hat{J}_{i,I}(t) |\0\>_{\mr{e}}
                        + \int_t^0 \d t' \, {}_{\mr{e}}\<\0| \hat{J}_j \hat{F}_{\l,I}(t') \hat{J}_{i,I}(t)  |\0\>_{\mr{e}}
                          \int_{-\infty}^t \d t' \, {}_{\mr{e}}\<\0| \hat{J}_j \hat{J}_{i,I}(t) \hat{F}_{\l,I}(t') |\0\>_{\mr{e}} \right]\bigg) \right|_{\Roset}.
  \end{split}
\end{equation}
\end{widetext}
Note that once the time ordering is fixed, we can use $\hat{J}_{j,I}(0) = \hat{J}_j$.
To evaluate the $t'$-integral, we make use of the identity
\begin{equation}
  \left. \hat{F}_{\l,I}(t') \right|_{\Roset}
  = -i \frac{\pd}{\pd t'} \left. \left[ \left( \frac{\pd}{\pd Q_{\l}} \mr{e}^{i \he t'} \right) \mr{e}^{-i \he t'} \right] \right|_{\Roset},
\end{equation}
which follows from straightforward algebraic manipulation.
We can then express the $t'$-integral in terms of the operator
\begin{equation}
  \hat{A}_{\l}(t') \equiv \left. \left[ \left( \frac{\pd}{\pd Q_{\l}} \mr{e}^{i \he t'} \right) \mr{e}^{-i \he t'} \right] \right|_{\Roset}.
\end{equation}
Noting that $\hat{A}_{\l}(0)=0$, we find:
\begin{equation}
  \begin{split}
      & \int_{-\infty}^{+\infty} \d t' \, \left. \left\{ {}_{\mr{e}}\<\0| \mathcal{T}\left[ \hat{F}_{\l,I}(t')
        \hat{J}_{i,I}(t) \hat{J}_{j,I}(0) \right] |\0\>_{\mr{e}} \right\} \right|_{\Roset} \\
    =& (-i) \bigg(   {}_{\mr{e}}\<\0| \hat{A}_{\l}(\infty) \mathcal{T} \left[ \hat{J}_{i,I}(t) \hat{J}_{j,I}(0) \right] |\0\>_{\mr{e}} \\
     &       \qquad - {}_{\mr{e}}\<\0| \mathcal{T} \left\{ \left[ \hat{A}_{\l}(t), \hat{J}_{i,I}(t) \right] \hat{J}_{j,I}(0) \right\} |\0\>_{\mr{e}} \\
     &       \qquad - {}_{\mr{e}}\<\0| \mathcal{T} \left[ \hat{J}_{i,I}(t) \hat{J}_{j,I}(0) \right] \hat{A}_{\l}(-\infty) |\0\>_{\mr{e}}
    \left. \bigg) \right|_{\Roset}.
  \end{split}
\end{equation}
The commutator in the third line can be rewritten as
\begin{equation}
    \left. \left[ \hat{A}_{\l}(t), \hat{J}_{i,I}(t) \right] \right|_{\Roset}
  = \left. \left[ \frac{\pd}{\pd Q_{\l}} \hat{J}_{i,I}(t) \right] \right|_{\Roset}
\end{equation}
\newline
and noting that $\hat{J}_{j,I}(0)=\hat{J}_j$ is independent of $\Rset$, we find
\begin{equation}
  \begin{split}
     & \left. \left( {}_{\mr{e}}\<\0| \mathcal{T} \left\{ \left[ \hat{A}_{\l}(t), \hat{J}_{i,I}(t) \right] \hat{J}_{j,I}(0) \right\}
       |\0\>_{\mr{e}} \right) \right|_{\Roset} \\
    =& \left. \left( {}_{\mr{e}}\<\0| \frac{\pd}{\pd Q_{\l}} \left\{ \mathcal{T} \left[ \hat{J}_{i,I}(t) \hat{J}_{j,I}(0) \right] \right\}
       |\0\>_{\mr{e}} \right) \right|_{\Roset}.
  \end{split}
\end{equation}
The other two terms, which involve the action of $\hat{A}_{\l}(\pm\infty)$
on the electronic part of the adiabatic ground state, can also be evaluated by straightforward algebraic manipulation
and by inserting a complete set of eigenstates of $\he(\Rset)$.
Note that we understand the limit $t'\to\pm\infty$ in the adiabatic sense, i.e., as $t'\to\pm\infty(1-i\eta)$,
where $\eta=0^+$ is again a positive infinitesimal.
We find
\begin{equation}
  \begin{split}
    & \left. \left( {}_{\mr{e}}\<\0| \hat{A}_{\l}(\infty) \right) \right|_{\Roset} \\
    =& -\bigg[ \frac{\pd}{\pd Q_{\l}} {}_{\mr{e}}\<\0|
               + \left( \frac{\pd}{\pd Q_{\l}} {}_{\mr{e}}\<\0| \right) |\0\>_{\mr{e}} {}_{\mr{e}}\<\0| \\
     & \qquad + \left. \lim_{t'\to\infty(1-i\eta)} i \frac{\pd \Eeo}{\pd Q_{\l}} t' \, {}_{\mr{e}}\<\0| \bigg] \right|_{\Roset}
  \end{split}
\end{equation}
and
\begin{equation}
  \begin{split}
    & \left. \left( \hat{A}_{\l}(-\infty) |\0\>_{\mr{e}} \right) \right|_{\Roset} \\
    =& \bigg[ \frac{\pd}{\pd Q_{\l}} |\0\>_{\mr{e}}
              - |\0\>_{\mr{e}} {}_{\mr{e}}\<\0| \left( \frac{\pd}{\pd Q_{\l}} |\0\>_{\mr{e}} \right) \\
     & \qquad + \left. \lim_{t'\to-\infty(1-i\eta)} i \frac{\pd \Eeo}{\pd Q_{\l}} t' \, |\0\>_{\mr{e}} \bigg] \right|_{\Roset}.
  \end{split}
\end{equation}
}

We can finally combine all individual pieces to obtain
\begin{widetext}
\begin{equation}
  \begin{split}
     &  \int_{-\infty}^{+\infty} \d t' \, \left. \left\{ {}_{\mr{e}}\<\0| \mathcal{T}\left[ \hat{F}_{\l,I}(t')
        \hat{J}_{i,I}(t) \hat{J}_{j,I}(0) \right] |\0\>_{\mr{e}} \right\} \right|_{\Roset} \\
    =& i \left. \left( \frac{\pd}{\pd Q_{\l}} \left\{ {}_{\mr{e}}\<\0| \mathcal{T} \left[
                       \hat{J}_{i,I}(t) \hat{J}_{j,I}(0) \right] |\0\>_{\mr{e}} \right\} \right) \right|_{\Roset} \\
     & + \left. \left( \left[   \lim_{t'\to\infty(1-i\eta)} \frac{\pd \Eeo}{\pd Q_{\l}} t'
                              - \lim_{t'\to-\infty(1-i\eta)} \frac{\pd \Eeo}{\pd Q_{\l}} t' \right]
         {}_{\mr{e}}\<\0| \mathcal{T} \left[ \hat{J}_{i,I}(t) \hat{J}_{j,I}(0) \right] |\0\>_{\mr{e}} \right) \right|_{\Roset},
  \end{split}
  \label{eq:deriv-jjcorfun}
\end{equation}
\end{widetext}
where we used the fact that $\pd/(\pd Q_{\l}) \, {}_{\mr{e}}\<\0|\0\>_{\mr{e}} = 0$ to simplify the expression.
Finally, we can use the basic identity
\begin{equation}
  \begin{split}
     &          \lim_{t'\to\infty(1-i\eta)} \frac{\pd \Eeo}{\pd Q_{\l}} t'
              - \lim_{t'\to-\infty(1-i\eta)} \frac{\pd \Eeo}{\pd Q_{\l}} t' \\
    =& \int_{-\infty}^{+\infty} \d t' \, {}_{\mr{e}}\<\0| \frac{\pd \he}{\pd Q_{\l}} |\0\>_{\mr{e}}
    =  \int_{-\infty}^{+\infty} \d t' \, {}_{\mr{e}}\<\0| \hat{F}_{\l,I}(t') |\0\>_{\mr{e}}
  \end{split}
\end{equation}
to identify the second term as the disconnected part of the correlation function,
which is exactly canceled by the second term in Eq.~(\ref{eq:ad-corfun-connect}).
Noting that all operators and states appearing in Eq.~(\ref{eq:deriv-jjcorfun}) are purely electronic ones,
we can let $\he\to\H^{\ad}$ and $|\0\>_{\mr{e}}\to|\0\>$ again
due to the additive and direct product nature of the adiabatic Hamiltonian and ground state, respectively.
We then finally arrive at
\begin{equation}
  \begin{split}
    &\sum_{I,a} \left(v^{\q=\mf{0},\l}_{I,a}\right)^*
    \left. \left[ \tilde{\cM}_{F_{I,a}}(\o'=0,\o;i,j) \right]\right|_{\mr{ad.}} \\
    & \qquad \qquad \qquad = i \left. \frac{\pd \tilde{\chi}^{\ad}_{ij}(\o)}{\pd Q_{\l}} \right|_{\Roset},
  \end{split}
\end{equation}
which completes the proof of Eq.~(\ref{eq:ad-limit-conjecture}).

We have thus shown that the full theory presented in this work reduces to the approach of static first derivatives
in the adiabatic limit, \emph{provided the $\o'$-dependence of the reduced matrix element is neglected as well.}
\newline


\section{Note on the practical treatment of two-phonon Raman scattering within the presented formalism}
\label{sec:app-twoph}

In this third Appendix, we want to briefly sketch how also the two-phonon-mediated part of the Raman spectrum
can be described in a practical way within the presented formalism.

Since the overlap $\<\a|\uu_{I,i}|0\>$ for two-phonon-like eigenstates of the matter Hamiltonian is rather small in typical cases,
being directly related to anharmonic effects and actually vanishing identically in the non-interacting limit,
it is far more practical to \emph{not} choose $\{\hat{O}_i\} = \{\uu_{I,i}\}$,
but rather choose $\{\hat{O}_i\} = \{\uu_{I,i}\uu_{J,j}\}$ instead.
The product of \emph{two} nuclei displacements typically has large overlap matrix elements $\<\a| \uu_{I,i}\uu_{J,j} |0\>$
for two-phonon-like eigenstates $|\a\>$ of the exact matter Hamiltonian.
The correlation function to consider is then
\begin{equation}
  S(t',t;\Lambda) = \<0| \mathcal{T} \left[ \uu_{I,i}(t') \uu_{J,j}(t') \hat{J}_{\kout,\n}(t) \hat{J}\dag_{\kin,\m}(0) \right] |0\>.
\end{equation}

To reduce this correlation function to a purely electronic one, one needs to consider both the first-order coupling
$\sum_{I,i} \uu_{I,i} \hat{F}_{I,i}$ in a second-order perturbative expansion as well as the second-order coupling
$\sum_{I,i;J,j} \uu_{I,i} \uu_{J,j} \hat{F}^{(2)}_{I,i;J,j}$ in a first-order expansion,
where $\hat{F}^{(2)}_{I,i;J,j}$ is the analogue of $\hat{F}_{I,i}$ (see Eq.~(\ref{eq:elecforce}))
involving the second derivatives of the electron-nuclei Coulomb interaction.
The relevant factorization approximation in this case is
\begin{widetext}
\begin{equation}
  \begin{split}
    S(t',t;\Lambda)
    \simeq& \sum_{K,k;L,l} \int \d t'' \int \d t''' \, \<0| \mathcal{T} \left[ \uu_{I,i}(t') \uu_{J,j}(t') \uu_{K,k}(t'') \uu_{L,l}(t''') \right] |0\> \\
    & \times \bigg\{ \<0| \mathcal{T} \left[ \hat{F}_{K,k}(t'') \hat{F}_{L,l}(t''') \hat{J}_{\kout,\n}(t) \hat{J}\dag_{\kin,\m}(0) \right] |0\>
     + \delta(t''-t''') \<0| \mathcal{T} \left[ \hat{F}^{(2)}_{K,k;L,l}(t'') \hat{J}_{\kout,\n}(t) \hat{J}\dag_{\kin,\m}(0) \right] |0\> \bigg\}.
  \end{split}
  \label{eq:twophfactor}
\end{equation}
\end{widetext}
As demonstrated in Section~\ref{sec:lsz}, the LSZ reduction formula then ``converts''
the exact two-phonon correlation function in the first line of Eq.~(\ref{eq:twophfactor})
into ``the square root'' of the exact two-phonon spectral function.
The final expression for the Raman scattering rate then involves the product of the exact two-phonon spectral function
and the square of a reduced matrix element involving only electronic operators,
given in the real-time domain by the second line of Eq.~(\ref{eq:twophfactor}).

Applying the procedure given in Appendix~\ref{sec:app-oneph} to the first of the two terms of this reduced matrix element
then leads to the independent-particle version of two-phonon Raman scattering theory
that has been used and discussed in the literature before~\cite{venezuela2011,herziger2014,reichardt2017}.
The influence of the second term involving the quadratic electron-nuclei/phonon coupling has so far not been discussed
and is currently work-in-progress, while excitonic effects can be captured by avoiding going to the independent-particle approximation
in the same way as in the case of one-phonon-mediated Raman scattering~\cite{reichardt2018c}.


%

\end{document}